\documentclass[preprint,showkeys]{revtex4-2}
\usepackage{graphicx}
\usepackage[colorlinks=true,urlcolor=blue,citecolor=blue,linkcolor=blue]{hyperref}
\usepackage{xcolor}

\begin{document}
	
	\title{Raman and first-principles study of the pressure induced Mott-insulator to metal transition in bulk FePS$_3$}
	
	\author{Subhadip Das$ ^{1} $}
	\thanks{Both authors contributed equally to this work.}
	\author{Shashank Chaturvedi$ ^{2,3} $}
	\thanks{Both authors contributed equally to this work.}
	\author{Debashis Tripathy$ ^{4} $}
	\author{Shivani Grover$^{2,3}$}
	\author{Rajendra Singh$^3$}
	\author{D. V. S. Muthu$ ^{1} $}
	\author{S. Sampath$ ^{4} $}
	\author{U. V. Waghmare$ ^{2} $}
	\author{A. K. Sood$ ^{1} $}
	\date{\today}
	\email{asood@iisc.ac.in}
	\affiliation{$ ^{1} $Department of Physics, Indian Institute of Science, Bangalore-560012, India\\
		$ ^{2} $Theoretical Sciences Unit, School of Advanced Materials, Jawaharlal Nehru Centre for Advanced Scientific Research, Bangalore-560064, India\\
		$ ^{3} $Chemistry and Physics of Materials Unit, School of Advanced Materials, Jawaharlal Nehru Centre for Advanced Scientific Research, Bangalore-560064 India\\
		$ ^{4} $Department of Inorganic and Physical Chemistry, Indian Institute of Science, Bangalore-560012, India}

	\begin{abstract}
		Recently discovered class of 2D materials based on transition metal phosphorous trichalcogenides exhibit antiferromagnetic ground state, with potential applications in spintronics. Amongst them, FePS$ _{3} $ is a Mott insulator with a band gap of $\sim$ 1.5 eV. This study using  Raman spectroscopy along with first-principles density functional theoretical analysis examines the stability of its structure and electronic properties under pressure. Raman spectroscopy reveals two phase transitions at 4.6 GPa and 12 GPa marked by the changes in pressure coefficients of the mode frequencies and the number of symmetry allowed modes. FePS$_3$ transforms from the ambient monoclinic C2/m phase with a band gap of 1.54 eV to another monoclinic C2/m (band gap of 0.1 eV) phase at 4.6 GPa, followed by another transition at 12 GPa to the metallic trigonal P-31m phase. Our work complements recently reported high pressure X-ray diffraction studies.
		
	\end{abstract}
	
	
	\maketitle
	\clearpage
	
	\section{Introduction}
\begin{figure}[b]
		\centering
		\includegraphics[width=\linewidth]{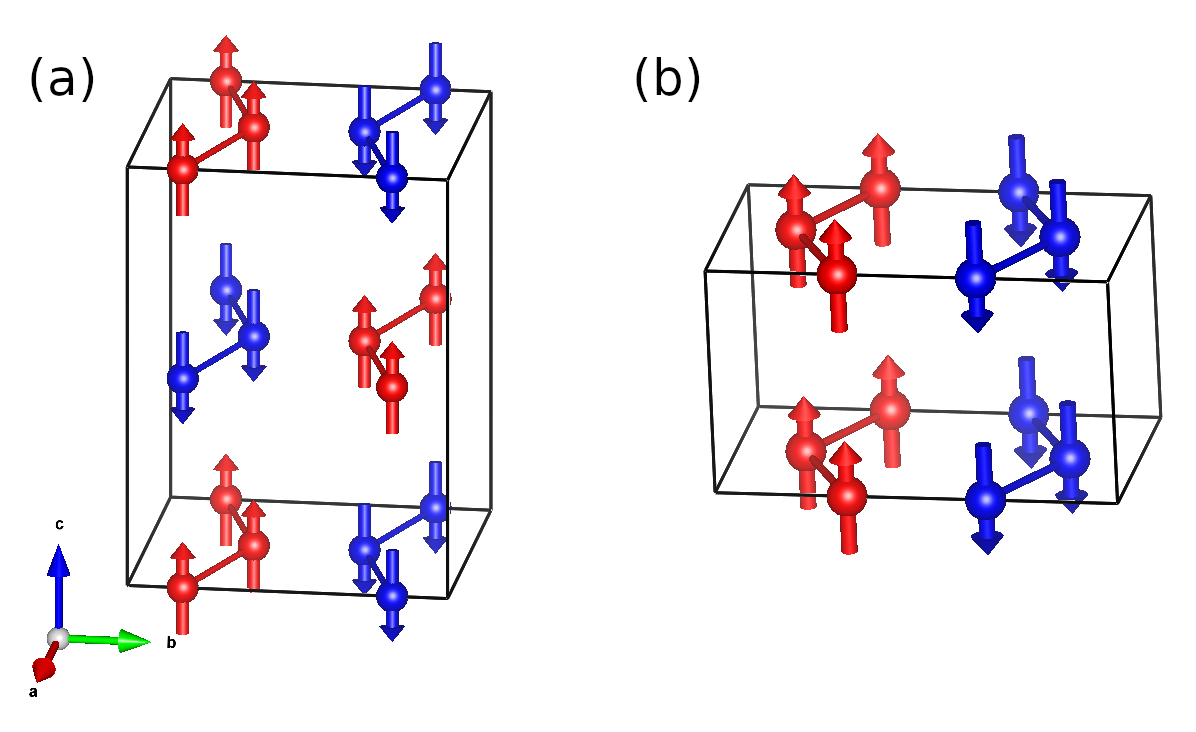}
		\caption{Magnetic ordering of $C2/m$ (a) LP (zAFM-I) and (b) HP-I (zAFM-II) phases \cite{PhysRevX.11.011024}. P and S atoms are removed and only Fe atoms are showed. The figures were made using VESTA software \cite{vesta}.}
		\label{mag_order}
	\end{figure}

	Transition metal phosphorus trichalcogenides (TMPX$ _{3} $, where TM = Fe, Mn, Ni or V, and X=S or Se) are a class of Mott insulators \cite{doi:10.1021/ic00209a028, 2002NCimR..25f...1G, ZHUKOV1996647, PhysRevLett.121.266801} known for their tremendous potential in the field of electrocatalysis \cite{doi:10.1021/acsenergylett.6b00184,doi:10.1021/acscatal.9b03180}, batteries \cite{FUJII2017370}, photocatalysis \cite{photo}, photodetector \cite{Ramos2021,C8TC05011B}, sensors \cite{doi:10.1021/acssensors.9b02064,C9TA03214B}, spintronics \cite{spin} and ultrathin magnetic devices \cite{magevan,Mag_clement,LEFLEM1982455,mag_joy,benard}. FePS$_{3} $ is a member of this family having a layered structure belonging to monoclinic crystal system with C2/m space group with a periodic unit cell containing two formula units \cite{BREC19863}. Sulfur atoms follow ABCABC stacking along $c$ axis and the octahedral voids are occupied by P and Fe atoms \cite{BREC19863}. Due to low cleavage energy associated with the layered material, FePS$ _{3} $ can be exfoliated down to a few layers by micro mechanical cleavage and liquid exfoliation \cite{doi:10.1021/acsenergylett.6b00184,Ramos2021}. At a N\'eel temperature of T$ _{N} \sim$118 K, the material undergoes a phase transition from a paramagnet to a two-dimensional (2D) Ising-type zig-zag antiferromagnet (zAFM), with high spin moments (S=2) of Fe$ ^{2+} $ atoms pointing perpendicular to the basal plane of the lattice \cite{afmtemp,neutrino_mag} (Fig. \ref{mag_order}(a)). Raman spectra below T$ _{N} $ show additional modes due to lowering  of  the symmetry in the magnetic phase \cite{wang_2016,afmtemp}. The presence  of these modes  down to a monolayer confirms the survival of AFM ordering in 2D system with slight lowering  in T$ _{N} $ due to weak interlayer van der Waals interaction \cite{wang_2016}.
	
	Pressure is an excellent tool to tune the structure, electronic \cite{saty_lifs,saty_pdps,PhysRevB.99.024111,qi2016superconductivity} and topological \cite{PhysRevLett.110.107401} properties  \cite{saty_phop,saty_pdps,PhysRevB.99.024111} of materials.  Although previous electronic bandstructure calculation of bulk FePS$ _{3} $ predicted its metallic nature \cite{cal_hashemi}, experiments  confirmed the material to be a Mott insulator with a band gap of $\sim$ 1.5 eV, lowest among  the TMPX$ _{3} $  family \cite{doi:10.1021/acsnano.5b05927}. Recent high pressure X-ray diffraction experiments at room temperature reveal two phase transitions at $\sim$ 4 GPa and $\sim$ 14 GPa, where the first transition is from ambient C2/m phase to another C2/m phase and the second transition to metallic trigonal P-31m phase \cite{PhysRevLett.121.266801}. Carrier drift velocity and thermal performance of these high-pressure phases are affected by phonons renormalization which can be analyzed in a non-invasive  manner by Raman spectroscopy. Furthermore, according to recent neutron-scattering measurements \cite{PhysRevX.11.011024}, the low pressure (LP) phase and the first high pressure (HP-I) phase of bulk FePS$ _{3} $ between 6 and 14 GPa have distinct low-temperature magnetic structures, zAFM-I and zAFM-II, respectively (Fig. \ref{mag_order}). Moreover, in the second high-pressure phase (HP-II) between 14 to 18 GPa, long-range magnetic ordering is lost \cite{PhysRevX.11.011024}. The relative stability, vibrational signatures and nature of these magnetically ordered structures need to be understood better. 
	
	In this paper, we report high pressure Raman experiments and density functional theory (DFT) based analysis of bulk FePS$ _{3} $. Raman experiments show two transitions at 4.6 GPa and 12 GPa, as seen by the pressure coefficients of the mode frequencies and significant changes in number of  modes, respectively. The linewidth of the phosphorus and sulfur dominated mode at $\sim$ 379 cm$ ^{-1} $ shows an anomalous decrease  with pressure attributed to the  buckling of sulfur atoms. Our DFT+$ U $ calculations of the recently reported magnetic structures of bulk FePS$_{3}$ \cite{PhysRevX.11.011024} show that HP-I phase has a narrow gap of $\sim$ 0.1 eV, explaining the experimentally observed broadening of the linewidth of some of the Raman modes with pressure and drop in resistivity above 5 GPa reported earlier \cite{PhysRevLett.121.266801}. The HP-II phase is an ordinary metal, without any long-range magnetic ordering. Comparison of pressure dependent frequencies of Raman modes determined experimentally with our calculations provides a clearer picture of the transitions.

	\section{Experimental details}
	High quality and large sized single crystals of FePS$ _{3} $ were synthesized by chemical vapor transport technique from the respective elements. Briefly, Fe, P and S powders were taken in 1:1:3 stoichiometric ratio and mixed well by grinding in an agate mortar. The mixture was transferred to a quartz tube (length 20 cm and inner diameter 18 mm) with iodine as the transport agent and sealed at a pressure of 2.7 x 10$ ^{-5} $ mbar. The tube was then transferred to a two-zone furnace with the hot zone kept at 900$^{\circ}$C and cold zone at 850$^{\circ}$C. After the reaction for two weeks, shiny black colored, centimeter size crystals were obtained, and the crystals were used for further characterization. X-ray diffraction patterns (see Fig. S1 of the supplementary information (SI)) were recorded using PAN Analytical X-ray BV diffractometer with Cu K$\alpha$ (1.5418 \AA) as the X-ray source.
	
	A thin crystalline platelet of FePS$_3$ (dimension $ \sim $ 100 $ \mu m$), a small ruby chip, and  16:3:1 volumetric solution of methanol, ethanol, and water respectively as the pressure transmitting medium (PTM) were placed together on an indented stainless steel metal gasket in a hole of $\sim$ 200 $\mu m$ diameter in a Mao-Bell type diamond-anvil cell (DAC). The doublet peaks of Ruby fluorescence were used for pressure calibration. Room temperature Raman spectra were recorded in back-scattering geometry using a commercial Horiba LabRam HR-800 spectrometer coupled with a Peltier cooled CCD as the detector. A solid-state laser of 532 nm wavelength was focused on the crystal using a 50$\times$ long working distance objective lens. The resolution of the captured spectra is $\sim$ 0.55 cm$ ^{-1} $. To avoid any unwanted heating or chemical reactions, the incident laser power was kept below 5 mW. 
	
	\section{Computational details}
	We studied pressure dependence of structure, electronic properties and phonons of FePS$_3$ using first-principles Density Functional Theory (DFT) as implemented in Quantum ESPRESSO software package \cite{qe}. In calculations of electronic and magnetic properties, we used DFT+$U$ method \cite{Hubbard} with an effective on-site Coulomb repulsion parameter $U$ = $2.5$ eV for Fe $3d$ orbitals as were used earlier \cite{abinitio}. The exchange-correlation energy of electrons was treated within a generalized gradient approximation and PBE functional \cite{Hua, Perdew}. Energy cut-off of 50 Ry was used in truncating basis sets to represent electronic wave functions and energy cut-off of 400 Ry was used for charge density. As FePS$_3$ belongs to the family of two-dimensional materials with weak interactions between layers stacked along $c$-axis, we included Grimme-D2 dispersion corrections to account for van der Waals (vdW) pairwise $1/r^6$ interactions \cite{grimme}. A uniform mesh of 6$\times$4$\times$3 $k$-points \cite{monkhorst} was used in sampling of Brillouin-zone (BZ) integrations in calculations of the LP phase with zAFM-I magnetic ordering (see Fig.~\ref{mag_order}(a)). Similarly, a $6\times4\times6$ mesh of $k$-points was used in sampling BZ of HP-I phase with zAFM-II magnetic ordering (see Fig.~\ref{mag_order}(b)). The magnetic ordering of HP-I phase was confirmed recently by Coak $et. al.$ using neutron-diffraction experiments \cite{PhysRevX.11.011024}. In calculations of the trigonal P-31m (HP-II) phase, $6\times6\times6$ mesh of $k$-points was used in sampling of BZ integrations. $\Gamma$-point phonon modes were obtained using frozen-phonon approach maintaining the magnetic ordering of reference LP and HP-I phases.
	
	\section{Results and discussion}
	\subsection{Experimental results}
\begin{figure}[t!]
		\centering
		\includegraphics[width=1\linewidth]{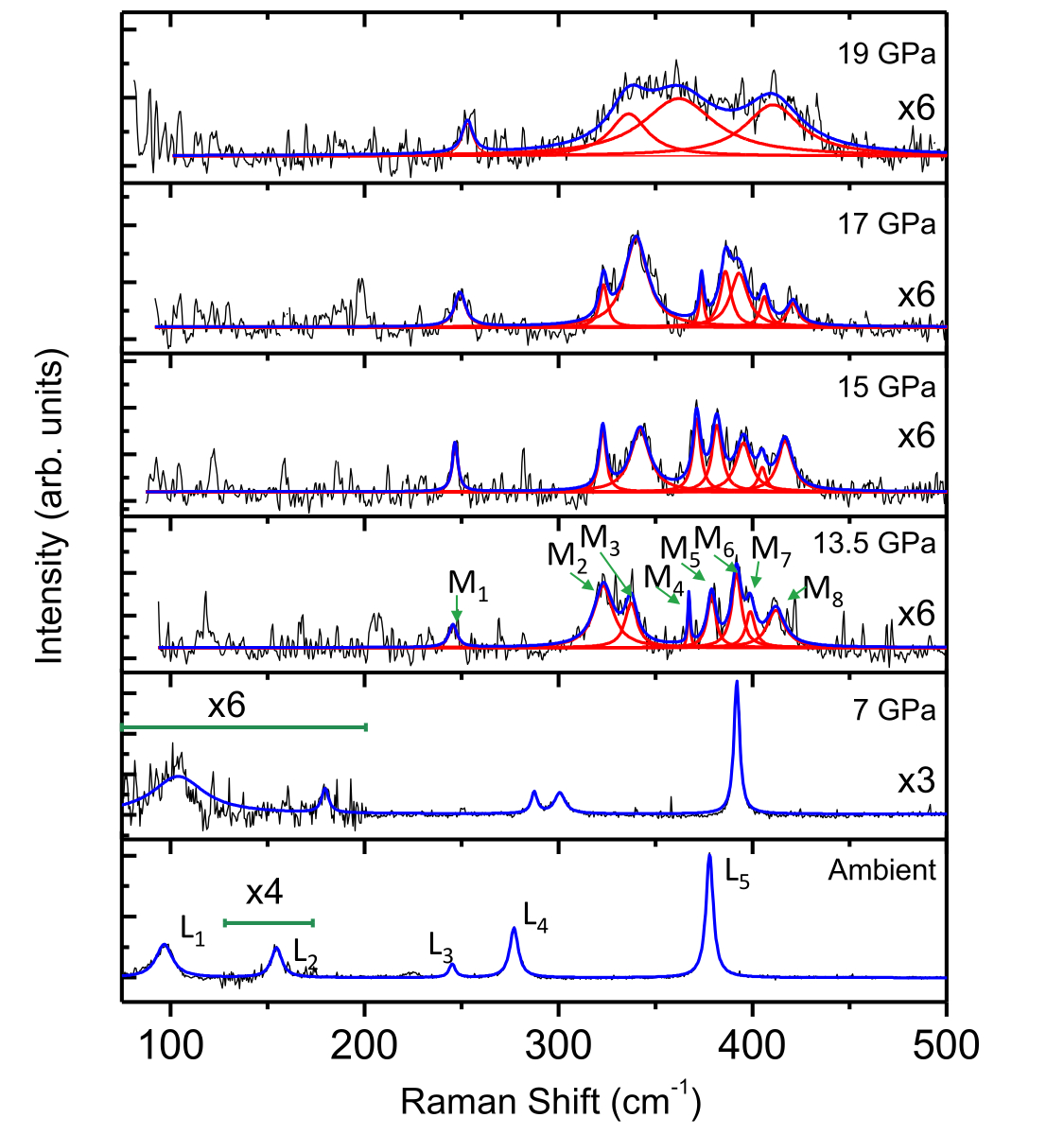}
		
		\caption{Evolution of the Raman modes of bulk FePS$ _{3} $ with pressure. Experimental data points are connected by black lines. Red and blue lines are their individual and cumulative Lorentzian peak fits, respectively. Modes are labeled in the figure. Intensities of some regions of the spectra are enhanced. }
		
		\label{f1}
	\end{figure}

	In the absence of magnetic ordering, bulk FePS$ _{3} $ at ambient conditions belongs to the C2/m space group, resulting in 30 zone center phonon modes ($ \Gamma=8A_{g}+6A_{u}+7B_{g}+9B_{u} $) \cite{wang_2016,doi:10.1021/acs.jpcc.7b09634}. Fig. \ref{f1} shows the  Raman modes of bulk FePS$ _{3} $ from 75 to 500 cm$ ^{-1} $ at ambient pressure. The modes are labeled from L$ _{1} $ to L$ _{5} $, where L$ _{3} $ and L$ _{5} $ modes are the A$ _{g}$-modes while L$ _{2} $ and L$ _{4} $ are combinations of the A$ _{g} $ and B$ _{g}$ modes \cite{wang_2016,PhysRevB.38.12089}. The L$ _{2} $ mode is iron dominated whereas L$ _{3} $, L$ _{4} $ and L$ _{5} $ modes originate from the molecular-like vibration of the [P$ _{2} $S$ _{6} $]$ ^{4-} $ unit \cite{wang_2016, afmtemp}. The broad and asymmetric mode L$ _{1} $  at $\sim$ 130 cm$ ^{-1} $ is reportedly iron-dominated  M-point phonon mode that gets resolved into multiple modes at T$ _{N}\sim$118 K from anti-ferromagnetic ordering of the Fe atoms \cite{afmtemp}. 
	
\begin{figure}[t!]
		\centering
		\includegraphics[width=0.8\linewidth]{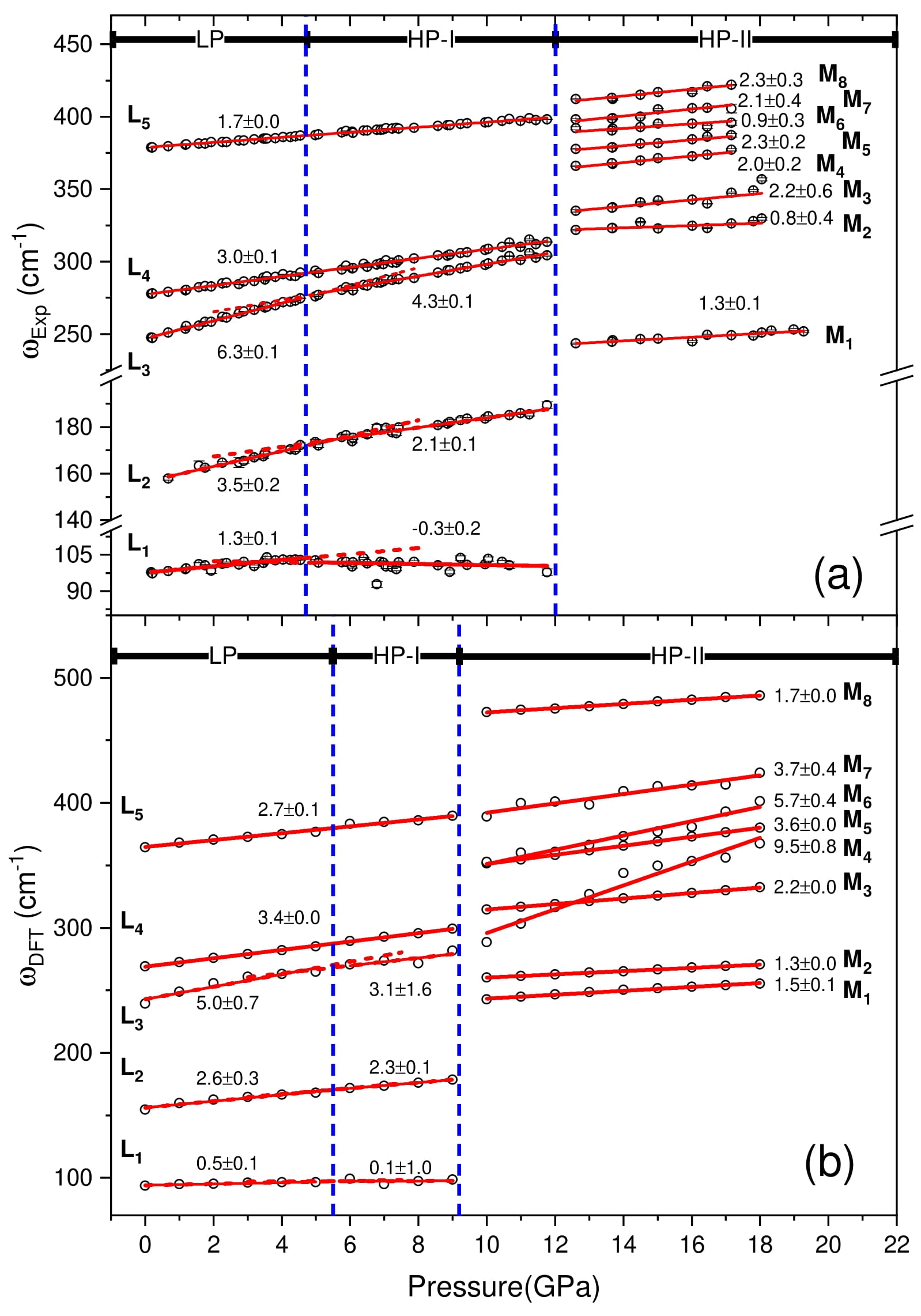}
		\caption {(a) Experimentally observed and (b) theoretically calculated pressure evolution of Raman modes. The slope of the red linear fit to the data  are shown in the figure. The dashed red lines for L$ _{1} $, L$ _{2} $ and L$ _{3} $ modes represent extensions to their linear fits. The three structural phases are indicated in the figure with the corresponding transition pressures highlighted by blue vertical dashed lines. } 
		
		\label{f2}
	\end{figure}

\begin{figure}[t!]
		\includegraphics[width=\linewidth]{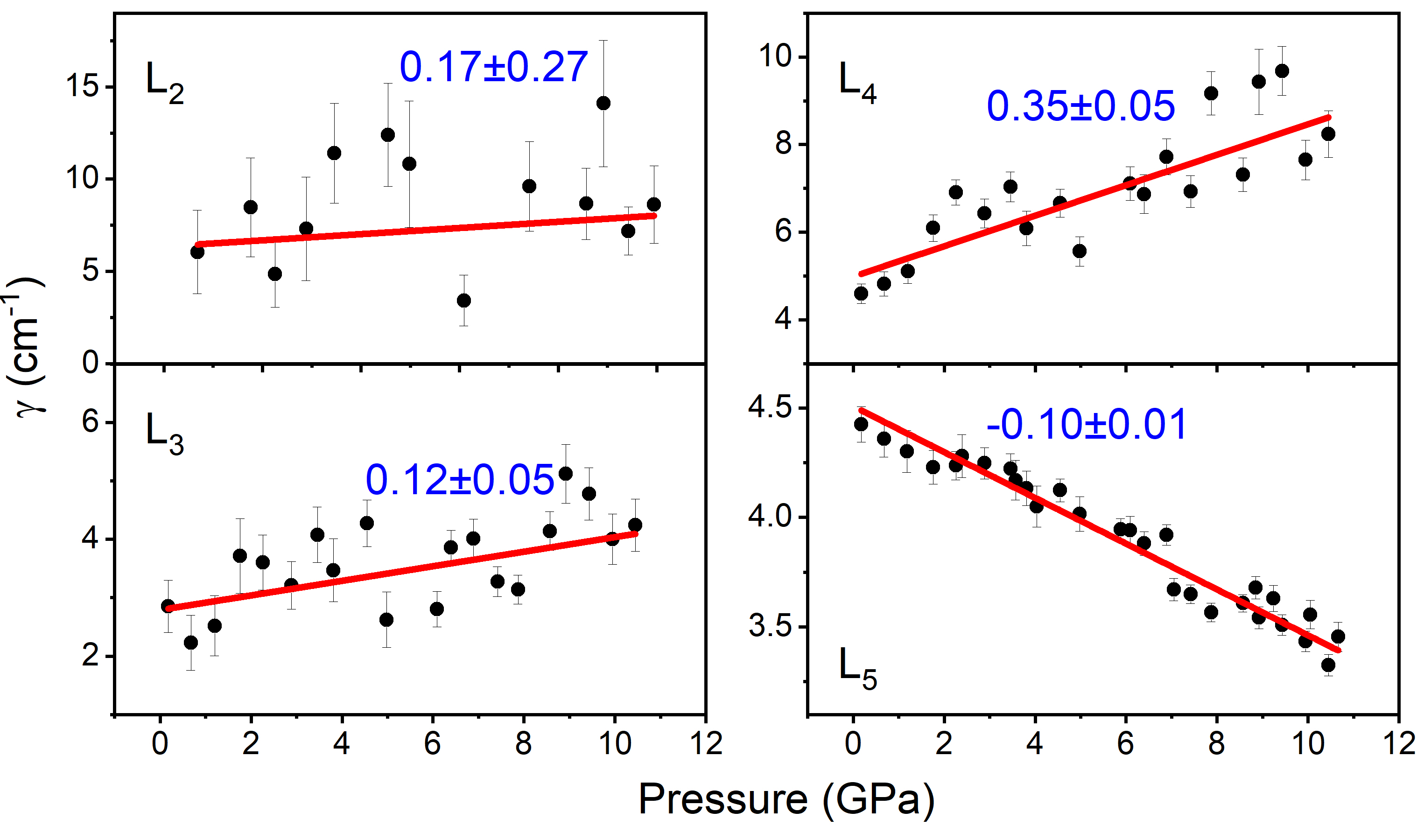}
		\caption {Linewidth ($\gamma$) of L$ _{2}$, L$ _{3}, $ L$ _{4} $ and L$ _{5} $ modes with pressure. Red lines are the linear fits. The slope is shown near the lines in cm$ ^{-1} $$\cdot$ GPa$ ^{-1} $ units.} 
		\label{f3}
	\end{figure}

	The evolution of Raman spectra at few representative pressures is shown in Fig. \ref{f1}. Above 12 GPa, the modes are labeled as M$ _{1} $ to M$ _{8} $. The spectra are fitted with a sum of Lorentzian functions to extract the pressure evolution of phonon frequencies ($\omega$) (Fig. \ref{f2}(a)) and linewidths ($\gamma$) (Fig. \ref{f3}). Although the Raman lineshape of the L$ _{1} $ mode is reported to be slightly asymmetric \cite{afmtemp}, we simply used a Lorentzian fitting function to get an estimate of the change in frequency with pressure. The modes L$ _{1} $ to L$ _{5} $ exhibit an expected linear increase in frequency with pressure, and are fitted with the linear equation, $ \omega(P)=\omega(P_{0})+C(P-P_0) $, where  $ \omega(P_{0}) $ is the frequency at ambient pressure and $ C $ is the pressure coefficient. These fitting parameters are noted in Fig. \ref{f2}(a) as well as summarized in Table-\ref{5t1}. Two transitions can be identified at P$ _{C1}\sim $ 4.6 GPa and P$ _{C2}\sim $ 12 GPa. We  observe a significant decrease of  pressure coefficients $ C $ for the L$ _{1} $, L$ _{2} $ and L$ _{3} $  modes above $\sim$ 4.6 GPa, with the number of Raman modes remaining the same. Notably, the iron dominated L$ _{1} $ mode shows a negative pressure coefficient in the HP-I phase.
	
	The first transition pressure P$ _{C1} $ is consistent with recent XRD measurements ($\sim$ 4 GPa) \cite{PhysRevLett.121.266801}, attributed to a change in the crystal angle $\beta$ from $\sim$ 107$^\circ $ to $\sim$ 90$^\circ $  and change in the atomic stacking of the iron and phosphorus atoms in the layer along the c$^ * $ axis with a higher inter-planar atomic coordination. Similar changes in the vibrational modes were also reported for V$ _{0.9} $PS$ _{3} $ at $\sim$ 12 GPa, which leads to an insulator to metal transition \cite{PhysRevB.100.035120}. We note that the L$ _{1} $ mode remains a single peak throughout the LP and HP-I phases, indicating the absence of any AFM ordering at room temperature \cite{afmtemp}.  
	
	As seen in Figs. \ref{f1} and \ref{f2}(a), Raman spectra changes considerably above P$ _{C2}\sim $ 12 GPa, marking the second phase transition. Above $\sim$ 12 GPa, eight distinct new modes are seen in the frequency range from 240 to 450 cm$ ^{-1} $. The  transition pressure (P$ _{C2} $) is very close to the  reported C2/m to P-31m phase transition at $ \sim $ 14 GPa from XRD measurements \cite{PhysRevLett.121.266801}. We note that the intensity of the Raman modes is reduced by almost half as compared to the HP-I phase. This is due to the metallic nature of the P-31m phase, as discussed in the next section. The frequencies of the Raman modes at 13 GPa along with their pressure coefficients ($ C $) are listed in Table-\ref{5t1}. As the modes M$ _{4} $ to M$ _{8} $ are close in frequency, we could analyze them till 17 GPa, after which they became too broad for further analysis.  
	
	The intrinsic linewidth of the phonon modes provides information about phonon-phonon, electron-phonon and spin-phonon interactions in a system  \cite{PhysRevLett.99.176802}. Thus in Fig. \ref{f3}, we have analyzed the linewidth of the Raman modes from L$ _{2} $ to L$ _{5} $ in the LP and HP-1 phases with pressure up to hydrostatic limit of our PTM ($\sim$ 10.5 GPa \cite{Klotz_2009}). Here we will not comment on the linewidth of the L$ _{1} $ mode, as it is reportedly asymmetric and cannot be deconvoluted at ambient temperature  \cite{afmtemp}. The linewidths of the L$ _{2} $, L$ _{3} $ and L$ _{4} $ modes increase with pressure. In the next section, we show that the HP-I phase is a narrow-gap semiconductor with a band gap of $\sim$ 0.1 GPa. Previous transport measurements also show that pressure induces significant reduction of resistivity above 5.5 GPa \cite{PhysRevLett.121.266801}. Thus, the electronic density of states (DOS) at the Fermi level increases with pressure, leading to an increase in the  EPC for the three modes. In contrast, the most intense L$ _{5} $ mode with dominant vibrations of phosphorus and sulfur atoms shows an anomalous decrease  in linewidth with pressure. This can be understood  from the reported pressure-induced buckling of the sulfur atoms due to a change in the stacking order along the c$ ^{*} $ axis \cite{PhysRevLett.121.266801}. The increased interaction between the sulfur atoms with pressure likely stabilizes the L$ _{5} $ mode, leading to the gradual decrease in linewidth. More theoretical work is needed to understand this effect. We make a note that for Bi$ _{2} $Te$ _{3} $ and Sb$ _{2} $Se$ _{3} $, similar anomalous Raman linewidths with pressure have been reported due to the electronic and structural transitions \cite{PhysRevLett.110.107401, https://doi.org/10.1002/pssb.201200672}.   
	
	\subsection{Results of theoretical analysis}
	\subsubsection{Transition pressures and electronic structure}
\begin{figure}
		\centering
		\includegraphics[width=0.5\textwidth]{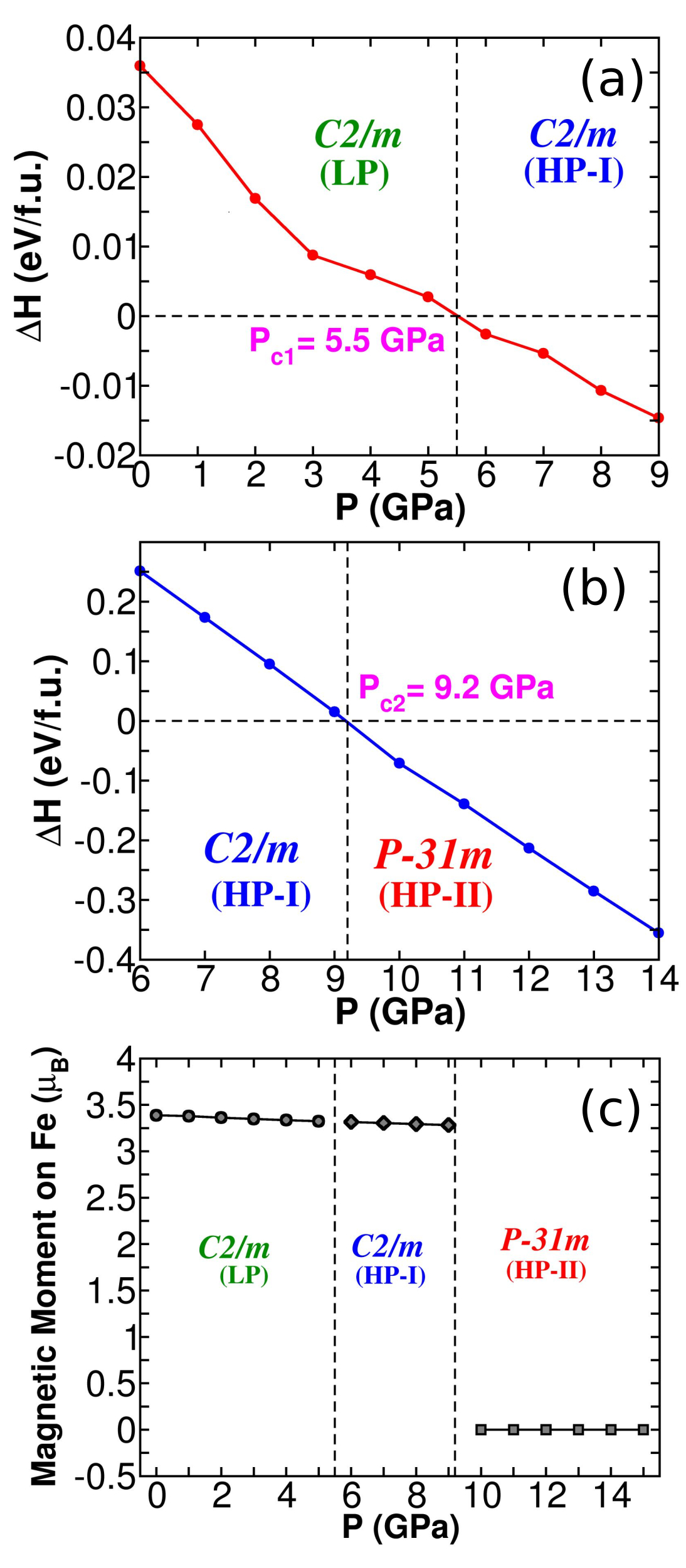}
		\caption{Enthalpy curves showing the evolution of crystal structure as function of pressure. (a) $P_{c1} = 5.5$ GPa corresponds to first structural transition from LP to HP-I phase and (b) $P_{c2} = 9.2$ GPa shows transition to high-pressure trigonal metallic phase. (c) Pressure dependence of calculated magnetic moments on Fe across phase transition. Calculations of the trigonal P-31m phase were performed on a non-magnetic state.}
		\label{enthalpy}
	\end{figure}

	We studied pressure dependent transitions with first-principles calculations of structures with appropriate magnetic ordering as reported experimentally using neutron diffraction \cite{flem,kurosawa,PhysRevX.11.011024}, shown in Fig. \ref{mag_order}. Earlier theoretical work considered zAFM-I magnetic ordering in their analysis of both the LP and HP-I phases \cite{abinitio,acs_nano}, which is not consistent with recent experiments showing zAFM-II magnetic ordering in the HP-I phase \cite{PhysRevX.11.011024}. Our calculations confirm the experiments that zAFM-I and zAFM-II are the magnetic ground states of the HP-I and HP-II phases respectively (see Table S1). Pressure dependence of the enthalpy difference (Fig. \ref{enthalpy}) of LP (zAFM-I) and HP-I (zAFM-II) phases shows that C2/m (HP-I) phase is stabilized at pressures above $P_{c1}=$ 5.5 GPa and P-31m (HP-II) phase becomes stable at pressures above $P_{c2}=$ 9.2 GPa. These transition pressures are quite close to the experimental estimates obtained with Raman spectra discussed in the previous section. Neutron scattering experiments have shown that HP-II phase has a short-range magnetic order, though precise magnetic ordering was not reported \cite{PhysRevX.11.011024}. This might explain a small underestimation of $P_{c2}$ in our calculations which assume non-magnetic ground state. Calculated lattice parameters are within the typical DFT errors (see Table S2) and in agreement with X-ray diffraction experiments \cite{PhysRevLett.121.266801}. The structural transitions at $P_{c1}$ and $P_{c2}$ are associated with decrease in volume and the $c$-lattice parameter (Fig. S2). Calculated estimate of the atomic moment on Fe at $P$ = 0 GPa is 3.4 $\mu_B$ (Fig. \ref{enthalpy}(c)), close to the reported value of 3.5 $\mu_B$ \cite{abinitio}.
	
\begin{figure}
		\includegraphics[width=0.8\linewidth]{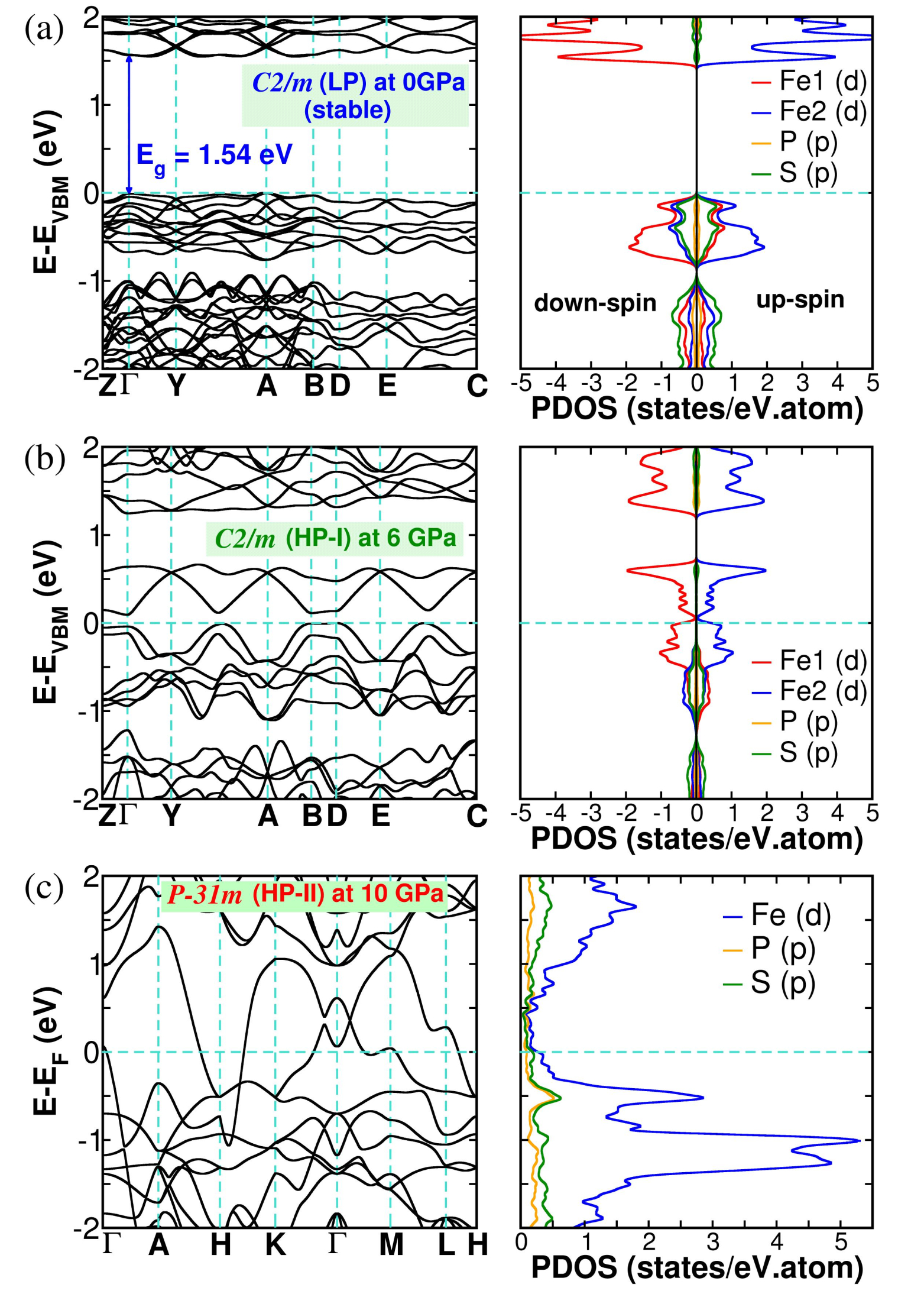}
		\caption{Electronic structure of LP, HP-I and HP-II phases of FePS$_3$. In LP phase a band gap of 1.54 eV is observed which reduced to $\sim$ 0.1 eV in HP-I phase stabilized under pressure. HP-II phase is metallic with finite contribution of Fe-$d$, P-$p$ and S-$p$ orbitals to the states at Fermi level.}
		\label{elec_struct}
	\end{figure}

	We find phonon instabilities at $\Gamma$-point of the LP phase, with most unstable mode at $153i$ cm$^{-1}$, comparable to recent calculations \cite{acs_nano} reporting $\Gamma$-point instability at 100$i$ cm$^{-1}$. The unstable modes of LP phase are shown in Fig. S3. Importantly, distorting the structure along eigenvectors of unstable mode at $153 i $ cm$^{-1}$ stabilizes the structure, lowering its energy by 8 meV/f.u. Electronic structure of these structures reveal no significant changes other than increase in the band gap from 1.32 eV to 1.54 eV (Fig. S6). We find no unstable phonon modes in HP-I and HP-II phases. From electronic structure of LP, HP-I and HP-II phases, it is clearly evident that the LP phase is an insulator, HP-I phase has a narrow band gap of $\sim$ 0.1 eV and HP-II phase is metallic (Fig. \ref{elec_struct}). Notable reduction in the band gap from 1.54 eV to 0.1 eV across the LP to HP-I phase transition explains the significant drop observed in resistance from $10^4$ $ \Omega$ at 4.5 GPa (LP phase) to $\sim 2$ $ \Omega$ at 5 GPa (HP-I phase) \cite{PhysRevLett.121.266801}. It is evident from electronic structure that pressure reduces band gap in HP-I phase ($E_g \sim$ 0.1 eV) and eventually drives it to metallic HP-II phase (Fig. \ref{elec_struct}). The projected density of states (PDOS) reveal hybridization of Fe-$d$ and P-$p$ orbitals with S-$p$ orbitals. In HP-II phase, there are contributions from Fe-$d$, P-$p$ and S-$p$ orbitals to the states at the Fermi level.
	
	\subsubsection{Calculated pressure dependence of Raman modes}
	We determined the pressure dependence of frequencies of phonons of stable LP, HP-I and HP-II phases, calculated using frozen-phonon method. The unstable LP (zAFM-I) and HP-I (zAFM-II) phases have space group C2/m, with 14 $A_g$ and 16 $B_g$ Raman modes respectively. The displacement of atoms along the maximum instability in LP phase breaks the symmetry of structure. This can be understood from the overlap of $A_g$ and $B_g$ phonon mode eigenvectors of unstable structure with the phonon mode eigenvectors of stable structure of LP phase (Fig. S9). This shows that structural distortion leads to non-zero overlap between multiple modes. In all the three phases, $c$-axis is the direction perpendicular to layers. Calculated mode frequencies as a function of pressure (Fig. \ref{f2}(b)) and their pressure coefficients ($C$) (Table \ref{5t1}) are close to the experimental values. In the LP phase, three low-frequency Raman active $B_g$ modes were observed (Fig. S5). The $B_g$ mode at 54 cm$^{-1}$ (Fig. S5(c)), involves displacements of individual monolayers against each other along $c$-axis. This mode hardens by $\sim$ 48 cm$^{-1}$ in going from 0 to 5 GPa, which is much larger than the changes in modes at 25 cm$^{-1}$ and 26 cm$^{-1}$ (Figs. S5(a) and S5(b)). The $B_g$ modes at 25 cm$^{-1}$ and 26 cm$^{-1}$, where monolayers slide against each other along $b$ and $a$-directions respectively, shows a weaker increase of $\sim$ 18 cm$^{-1}$ from 0 to 5 GPa (Fig. S5(a)). However, such increase in mode frequencies tapers off with pressure due to reduced vdW gap between the layers at high pressures. The $L_1$ mode of the LP and HP-I phases exhibits relatively low value of $C$ as they involve out-of-plane vibrations of Fe atoms only (Fig. S4(a) and S7(a)), and are not significantly affected by change in the vdW gap. The $L_3$ mode in the LP phase has highest $C$ $\sim$ 5 cm$^{-1}$GPa$^{-1}$ which reduces to $\sim$ 3 cm$^{-1}$GPa$^{-1}$ in HP-I phase. This is understandable as $L_3$ mode involves vibrations of only S atoms along the layered direction. Both $L_2$ and $L_3$ modes exhibit a change in pressure coefficient above $P_{c1}$. $L_4$ and $L_5$ modes did not undergo any change in $C$ across $P_{c1}$ transition (Table \ref{5t1}). The $L_2$, $L_4$ and $L_5$ modes of the LP and HP-I phases involve displacements of P and S atoms and hence exhibit higher $C$ than $L_1$ mode with Fe vibrations but lower than that of $L_3$ mode.
	
	The HP-II phase has 8 Raman active modes (5 $E_g$+3 $A_{1g}$). Comparison of experimental and calculated frequencies at 13 GPa shows that some of the observed modes cannot be assigned to the calculated frequencies (Table \ref{5t1}). We note that our calculated phonons at 18 GPa are in agreement with an earlier theoretical work (Table S3) \cite{jctc}. $E_g$ modes involve in-plane vibrations of Fe, P and S atoms, while $A_{1g}$ modes involve vibrations of P and S atoms against each other along $c$-direction (Fig. S8). From Fig. \ref{f2}(b), it is evident that the pressure coefficient of the $E_g$ modes in HP-II phase is weaker than that of the $A_{1g}$ modes. 
	
	\section{Conclusions}
	In summary, we have analyzed the electronic and vibrational properties of bulk FePS$ _{3} $ as a function of pressure through  Raman spectroscopy and first-principles calculations. We observe a structural phase transition at $P_c\sim$ 4.6 GPa, marked by reduction in the pressure coefficients of the Raman modes near 98 cm$ ^{-1} $, 158 cm$ ^{-1} $ and 248 cm$ ^{-1} $. This is accompanied by broadening of linewidth of $L_2$, $L_3$ and $L_4$ Raman modes till 10.5 GPa due to an increase in the EPC. In contrast, the sulfur and phosphorus dominated mode $L_5$ exhibits anomalous decrease of linewidth with pressure due to the previously reported buckling of the sulfur atoms. Eight new Raman modes were observed between the frequency range of 240 to 412 cm$ ^{-1} $ in the P-31m phase above 12 GPa. Our DFT+$ U $ calculations show that the LP phase is an insulator with an electronic gap of 1.54 eV which reduces drastically to $\sim$ 0.1 eV in HP-I phase and the HP-II phase is metallic. Our calculated values of transition pressures are close to the estimates obtained from Raman and X-ray diffraction experiments. Calculated frequencies and pressure coefficients of the Raman modes of LP and HP-I monoclinic phases qualitatively agree with our experiments. However, exact knowledge of experimentally reported short-range magnetic ordering in HP-II trigonal phase is needed to explain in detail the nature of its phonon modes and to get better estimate of $P_{c2}$. Symmetry analysis of the phonon eigenvectors reveals that the L$_3$ mode in the LP phase and A$ _{1g} $ modes in the HP-II phase have relatively higher pressure coefficients. These modes involve vibrations of P and S atoms along the layered direction. We hope our work may enable future applications of FePS$ _{3} $ and other members of TMPX$ _{3} $ compounds in ultrathin novel magnetic devices.  
	
	\section*{Acknowledgments} 
	A.K.S. thanks Department of Science and Technology, India for support under the Year of Science Professorship and Nanomission Council. S.C. acknowledges JNCASR for research fellowship and Thematic Unit of Excellence on Computational Material Science, JNCASR for computational resources. U.V.W. acknowledges support from a J.C. Bose National fellowship of SERB-DST, Govt. of India.
	
	\begin{table}
		\centering
		\caption{Comparison of the experimentally observed and theoretically calculated phonon frequency ($\omega$) and pressure coefficient ($ C $) of the Raman active modes at three structural phases of FePS$ _{3} $. Change in $ C $ are observed for L$ _{1} $, L$ _{2} $ and L$ _{3} $ modes across LP and HP-I phases.}
		\begin{tabular}{|c|c|c|c|c|c|c|}
			\hline
			Structure& Modes& Symmetry& \multicolumn{2}{c|}{Frequency (cm$ ^{-1} $)}& \multicolumn{2}{c|}{Pressure coefficient (cm$ ^{-1} $GPa$ ^{-1} $)}\\ \hline
			& & & $\omega_{Exp}$ (0 GPa)& $\omega_{DFT}$ (0 GPa) &  $ C_{Exp} $ & $ C_{DFT} $\\ \hline
			C2/m (LP)& L$ _{1} $& A$ _{g}$, B$ _{g} $ &97.8$\pm$ 0.4 &94 & 1.3$\pm$ 0.1& 0.5 $\pm$ 0.1  \\  
			& L$ _{2} $& A$ _{g} $& 158.5$\pm$0.6 &154 &3.5$\pm$ 0.2 &2.6 $\pm$ 0.3\\ 
			& L$ _{3} $& A$ _{g}$ &247.7$\pm$0.3 &239 &6.3$\pm$ 0.1& 5.0 $\pm$ 0.7\\ 
			& L$ _{4} $& A$ _{g}$, B$ _{g} $&278.0$\pm$0.2 &269 & 3.0$\pm$ 0.1& 3.4 $\pm$ 0.0\\ 
			& L$ _{5} $& A$ _{g}$& 378.6$\pm$0.2 &365 & 1.7$\pm$ 0.0& 2.7 $\pm$ 0.1\\ \hline  
			& & & $\omega_{Exp}$ (6 GPa)& $\omega_{DFT}$ (6 GPa)& & \\ \hline
			
			C2/m (HP-I)& L$ _{1} $& B$ _{g} $ &100.1$\pm$ 0.7 &99 & -0.3$\pm$ 0.2&0.1 $\pm$ 1.0 \\  
			& L$ _{2} $& B$ _{g} $& 173.8$\pm$0.6 &172 &2.1$\pm$ 0.1&2.3 $\pm$ 0.1\\ 
			& L$ _{3} $& A$ _{g}$ &281.5$\pm$0.2 &271 &4.3$\pm$ 0.1&3.1 $\pm$ 1.6\\ 
			& L$ _{4} $& B$ _{g}$&295.3$\pm$0.2 &290 & 3.0$\pm$ 0.1& 3.4 $\pm$ 0.0\\ 
			& L$ _{5} $& A$ _{g}$& 389.7$\pm$0.0 &383 & 1.7$\pm$ 0.0& 2.7 $\pm$ 0.1\\ \hline  
			& & & $\omega_{Exp}$ (13 GPa)& $\omega_{DFT}$ (13 GPa)&  & \\ \hline
			
			P-31m (HP-II) & M$ _{1} $ & E$ _{g}$ & 248.5 $\pm$ 2.3 & 259 & 1.3$\pm$0.1 & 1.5$\pm$0.1\\ 
			& M$ _{2} $  & E$ _{g}$ & 322.2 $\pm$ 6.4 & 264 & 0.8$\pm$0.4 &1.3$\pm$0.0  \\ 
			& M$ _{3} $  & E$ _{g}$& 335.2 $\pm$ 9.6 & 321& 2.2$\pm$0.6 &2.2$\pm$0.0  \\
			& M$ _{4} $  & A$ _{1g}$& 365.2 $\pm$ 2.6 & 327 & 2.3$\pm$0.2 &9.5$\pm$0.8  \\ 
			& M$ _{5} $  & E$ _{g}$ & 377.3 $\pm$ 3.3 & 362 & 2.0$\pm$0.2 & 3.6$\pm$0.0  \\ 
			& M$ _{6} $  & A$ _{1g}$ & 389.8 $\pm$ 5.2 & 366 & 1.6$\pm$0.4 & 5.7$\pm$0.4 \\
			& M$ _{7} $  & A$ _{1g}$& 397.2 $\pm$ 4.7 & 399& 2.1$\pm$0.3 & 3.7$\pm$0.4 \\
			& M$ _{8} $  & E$ _{g}$ & 411.2 $\pm$ 2.8 & 477 & 2.3$\pm$0.2 &1.7$\pm$0.0  \\\hline
			
		\end{tabular}
		\label{5t1}
	\end{table}

	\clearpage
	\bibliography{ref} 
	\bibliographystyle{apsrev4-2}

\end{document}


\title{Supplementary Information : Raman and first-principles study of the pressure induced Mott-insulator to metal transition in bulk FePS$_3$}

\author{Subhadip Das$ ^{1} $}
\thanks{Both authors contributed equally to this work.}
\author{Shashank Chaturvedi$ ^{2,3} $}
\thanks{Both authors contributed equally to this work.}
\author{Debashis Tripathy$ ^{4} $}
\author{Shivani Grover$^{2,3}$}
\author{Rajendra Singh$^3$}
\author{D. V. S. Muthu$ ^{1} $}
\author{S. Sampath$ ^{4} $}
\author{U. V. Waghmare$ ^{2} $}
\author{A. K. Sood$ ^{1} $}
\date{\today}
\email{asood@iisc.ac.in}
\affiliation{$ ^{1} $Department of Physics, Indian Institute of Science, Bangalore-560012, India\\
	$ ^{2} $Theoretical Sciences Unit, School of Advanced Materials, Jawaharlal Nehru Centre for Advanced Scientific Research, Bangalore-560064, India\\
	$ ^{3} $Chemistry and Physics of Materials Unit, School of Advanced Materials, Jawaharlal Nehru Centre for Advanced Scientific Research, Bangalore-560064, India\\
	$ ^{4} $Department of Inorganic and Physical Chemistry, Indian Institute of Science, Bangalore-560012, India}

\keywords{Raman spectroscopy, high-pressure study, Mott insulator, iron phosphorus trisulfide, insulator-metal transition, magnetic ordering.}

\maketitle

\clearpage
\section{Characterization}
\begin{figure}[H]
\centering
	\includegraphics[width=0.8\linewidth]{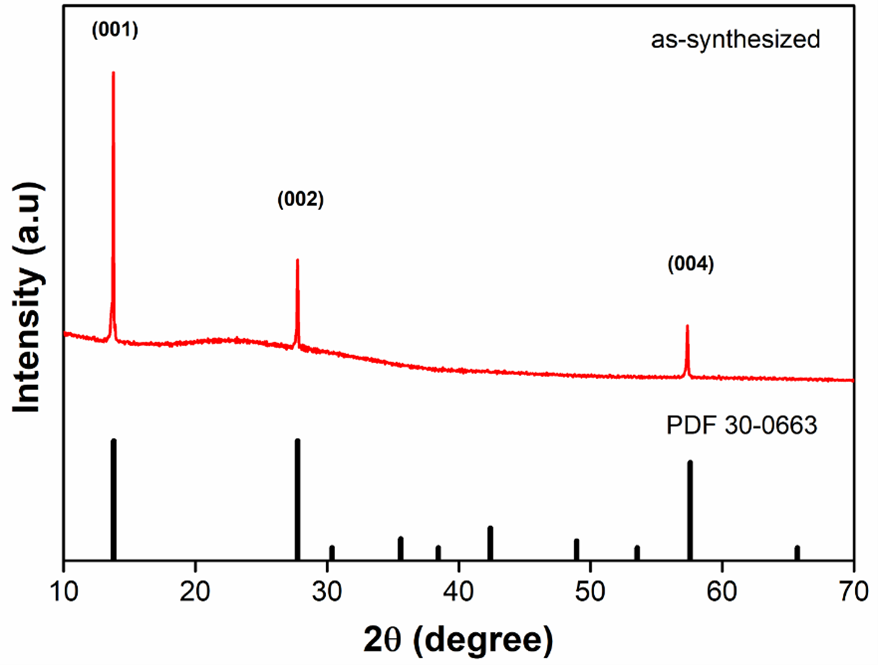}
	\caption{X-ray diffraction pattern of as-synthesized FePS$ _{3} $ crystals. The standard pattern is also given.}
	\label{xrd}
\end{figure}

Fig. \ref{xrd} shows the X-ray diffraction pattern of the as-synthesized FePS$ _{3} $ crystals. The crystal is oriented in $ (00l) $ direction suggesting the formation of pure, highly ordered crystals.

\clearpage

\section{DFT+$U$ calculations}

\begin{table}[H]
\centering
\caption{\small{Calculated energy differences of zAFM-I and zAFM-II magnetic ordering for both LP and HP-I phases.}}
    \begin{center}
    \begin{tabular}{|c|c|c|}
    \hline
& $C2/m$ (LP)& $C2/m$ (HP-I)\\
\hline
zAFM-I &Ground State & 31 meV/f.u\\
zAFM-II &2 meV/f.u.& Ground State\\
\hline

    \end{tabular}
    \end{center}
    \label{gs}
\end{table}

\begin{table}[H]
\centering
\caption{\small{Calculated value of lattice parameters of LP phase and HP-I for magnetic ordering as reported by Coak \textit{et. al.} \cite{PRX}, and for HP-II phase. Lattice parameters are in agreement with previous experimental report \cite{prl}.}}
    \begin{tabular}{|c|c|c|c|}
    \hline
Structural Parameter & $C2/m$ \newline (LP phase) & $C2/m$ \newline (HP-I phase) & P-31m \newline (HP-II phase)\\
 & 0 GPa & 9 GPa & 18 GPa\\
\hline
$a$ (\AA)&5.95&5.71&5.72\\
$b$ (\AA)&10.25&10.02 &5.72\\
$c$ (\AA)&6.76&5.82 &4.63\\
$\alpha^o$&90&90&90\\
$\beta^o$&107.26&91.07&90\\
$\gamma^o$&90&90&120\\
$V$ (\AA$^3$)&393.8&332.83&131.64\\
\hline

    \end{tabular}
    \label{latpar_tab}
\end{table}

\begin{table}[H]
\centering
\caption{\small{Comparison of the  phonon frequencies of the P-31m phase from our experimental measurements and previous calculations \cite{https://doi.org/10.1002/jcc.26178}.}}
 
 \begin{center}
    \begin{tabularx}{\textwidth}{>{\hsize=.6\hsize\linewidth=\hsize}X
                                                     >{\hsize=0.9\hsize\linewidth=\hsize}X
                                                     >{\hsize=0.9\hsize\linewidth=\hsize}X
                                                     >{\hsize=0.9\hsize\linewidth=\hsize}X 
                                                     >{\hsize=0.9\hsize\linewidth=\hsize}X 
                                                     >{\hsize=0.95\hsize\linewidth=\hsize}X}
    \hline \hline \\[-1ex]
			Mode & $\omega_{exp}$ (cm$ ^{-1} $) (at 12.6 GPa)& $\omega_{DFT}$ (cm$ ^{-1} $) (at 12 GPa) &$\omega_{DFT}$ (cm$ ^{-1} $) \newline(at 18 GPa) &$\omega_{DFT}$ (cm$ ^{-1} $) (at 18 GPa) from Ref. \cite{https://doi.org/10.1002/jcc.26178}  \\ 
			\hline \hline
			M$ _{1} $ & 243.4 $\pm$ 2.3 &  247 (E$ _{g}$)& 259 (E$ _{g}$) &259 (E$ _{g}) $ \\ 
			\hline
			                   &                              & 263 (E$ _{g}$) & 271 (E$ _{g}$) & 276 (E$ _{g}) $ \\
			 \hline
			M$ _{2} $ & 322.2 $\pm$ 6.4  & 317 (A$_{1g}$)  & 366 (A$_{1g}$)& 334 (A$ _{1g}) $\\ 
			                   &                              & 319 (E$ _{g}$) & 332 (E$ _{g}$) & 352 (E$ _{g}) $ \\
			                   \hline
			M$ _{3} $ & 335.2 $\pm$ 9.6 &  & & \\ 
			\hline
			M$ _{4} $ & 365.2 $\pm$ 2.6   &358 (E$ _{g}$) & 380 (E$ _{g}$) &380 (E$ _{g}) $ \\ 
			                   &                                &361 (A$ _{1g}$)&399 (A$ _{1g}$)&381 (A$ _{1g}) $  \\
			                   \hline
			M$ _{5} $ & 377.3 $\pm$ 3.3   &  & &\\ 
			\hline
			M$ _{6} $ & 389.8 $\pm$ 5.2  &  & &\\ 
			\hline
			M$ _{7} $ & 397.2 $\pm$ 4.7  &404 (A$ _{1g})$ & 417(A$ _{1g}) $  &416 (A$ _{1g}) $ \\ 
			\hline
			M$ _{8} $ & 411.2 $\pm$ 2.8 &  & & \\
			\hline
                     &  & 475 (E$_{g}$) & 486 (E$_{g}$) &486 (E$ _{g}) $ \\
                     \hline \hline
    \end{tabularx}
    \end{center}
    \label{t2}
\end{table}
\clearpage
\begin{figure}[H]
\centering
         \centering
         \includegraphics[width=0.45\textwidth]{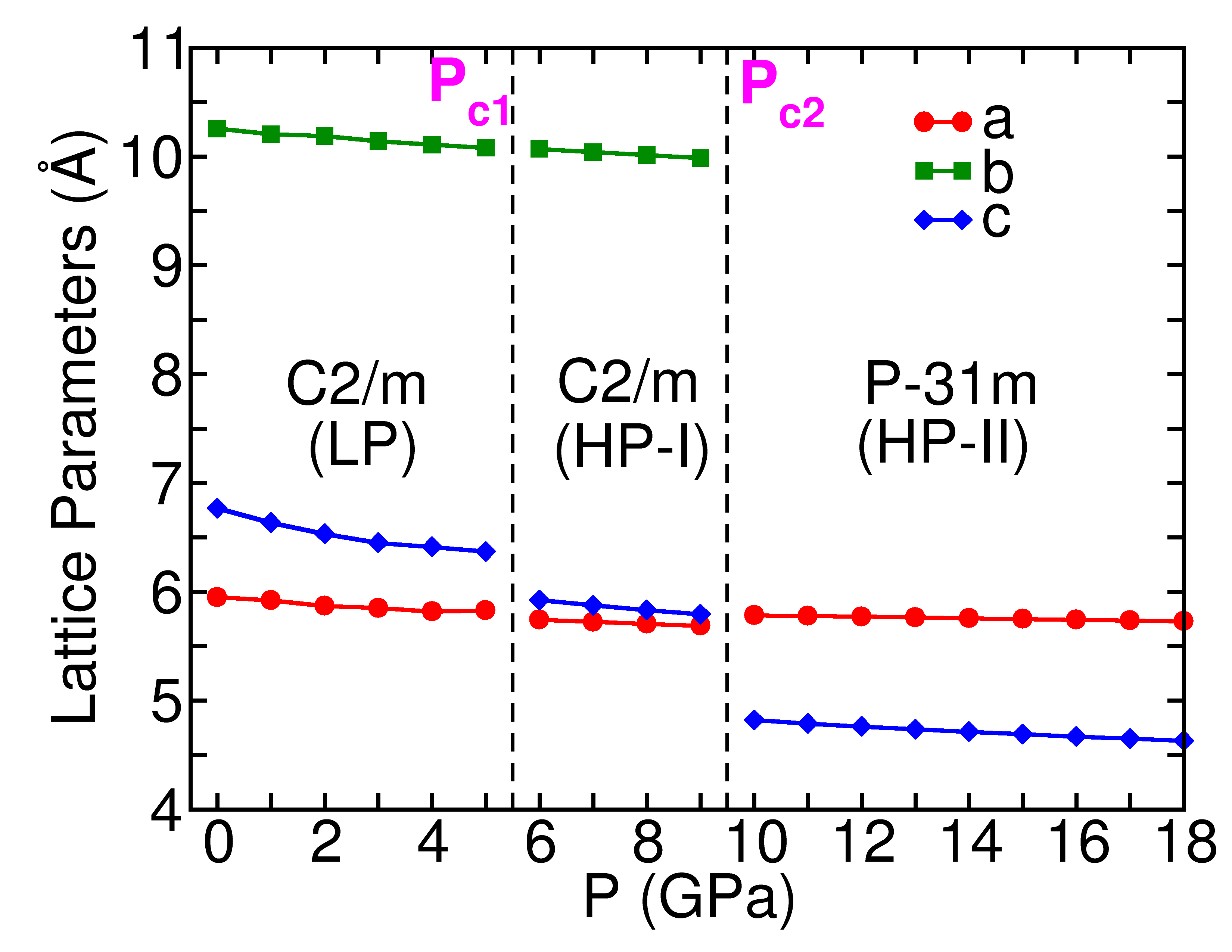}
         \includegraphics[width=0.45\textwidth]{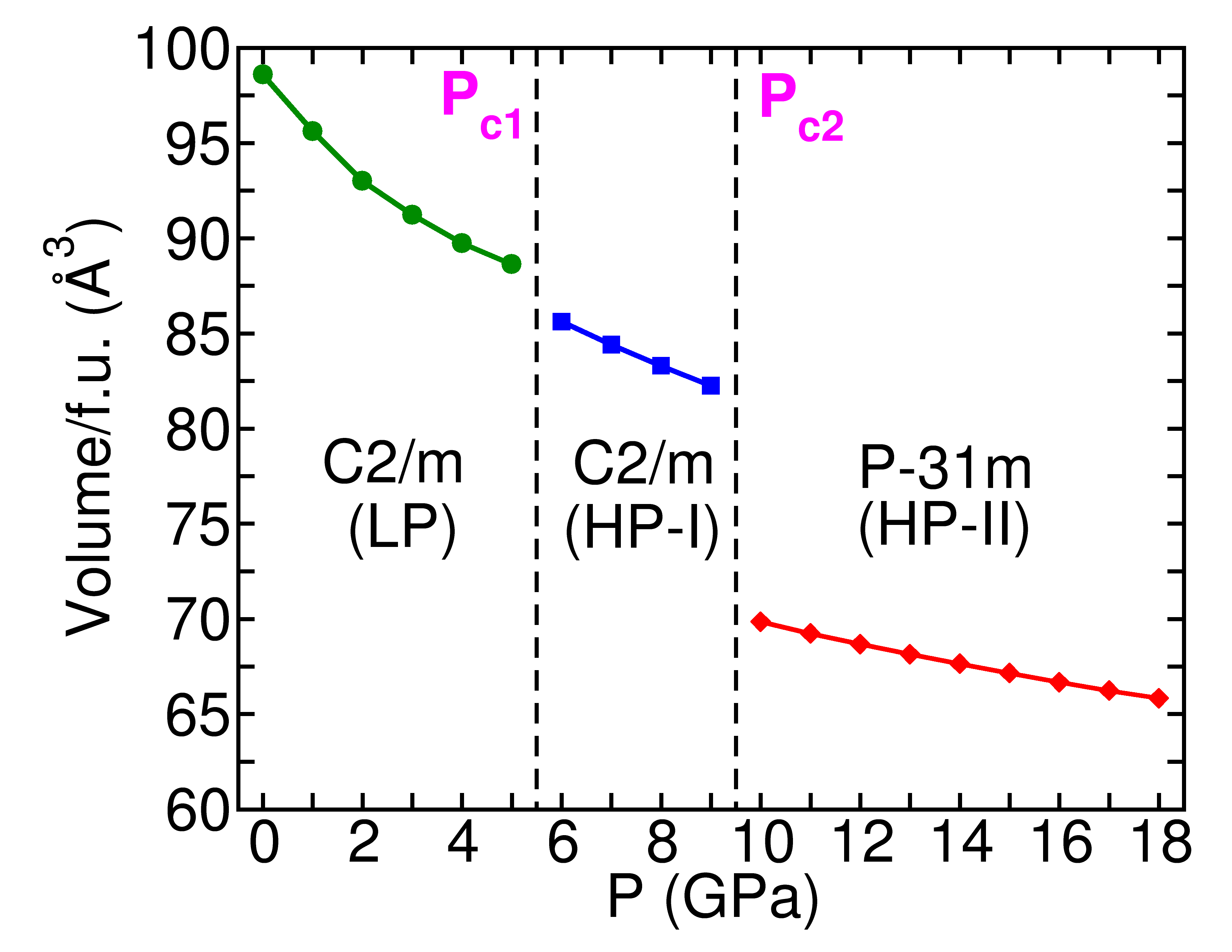}
        \caption{\small{Evolution of lattice parameters with pressure from first-principles calculations. The dotted line corresponds to theoretically determined transition pressures. Drop in volume across transition pressures is observed.}}
        \label{lat_par}
\end{figure}

\begin{figure}[H]
\centering
     \begin{subfigure}[b]{0.42\textwidth}
         \centering
         \includegraphics[width=\textwidth]{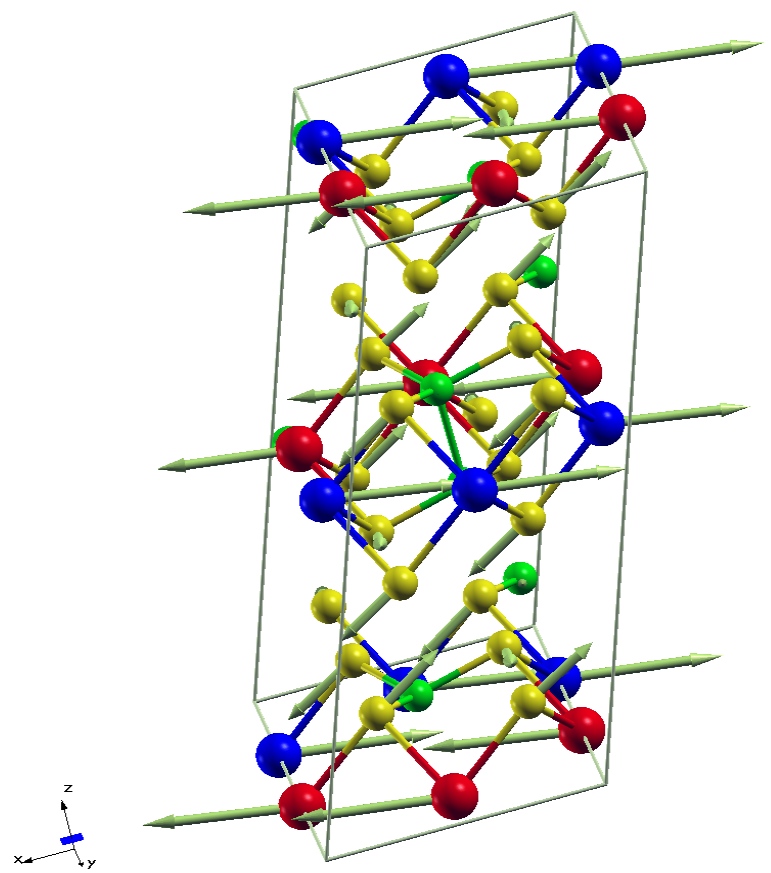}
        \caption{}
         \label{unstable1}
     \end{subfigure}
     \hfill
     \begin{subfigure}[b]{0.42\textwidth}
         \centering
        \includegraphics[width=\textwidth]{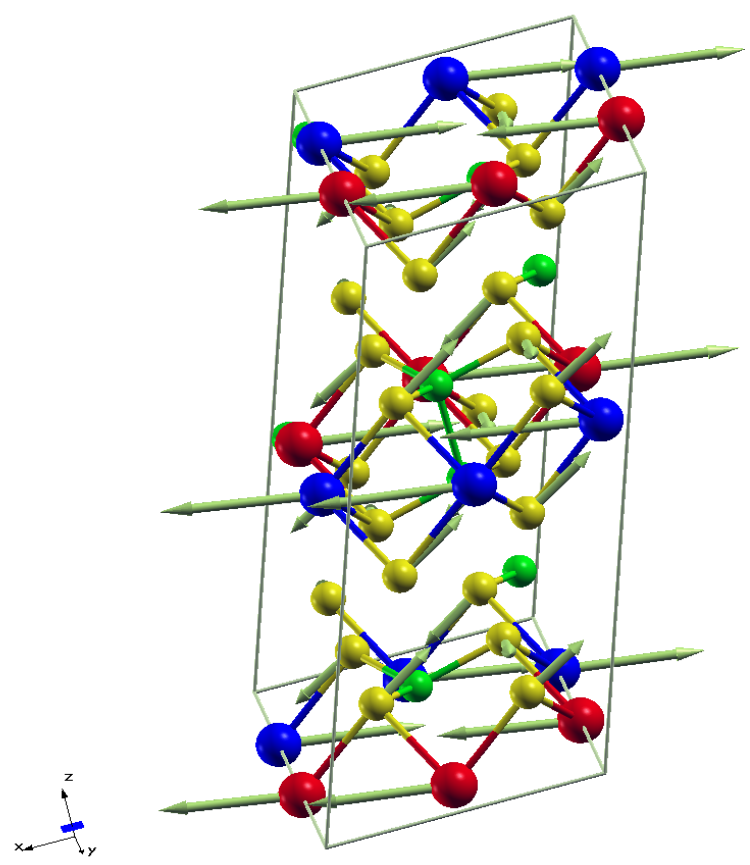}
         \caption{}
         \label{unstable2}
     \end{subfigure}
     \hfill
     \begin{subfigure}[b]{0.4\textwidth}
        \centering
         \includegraphics[width=\textwidth]{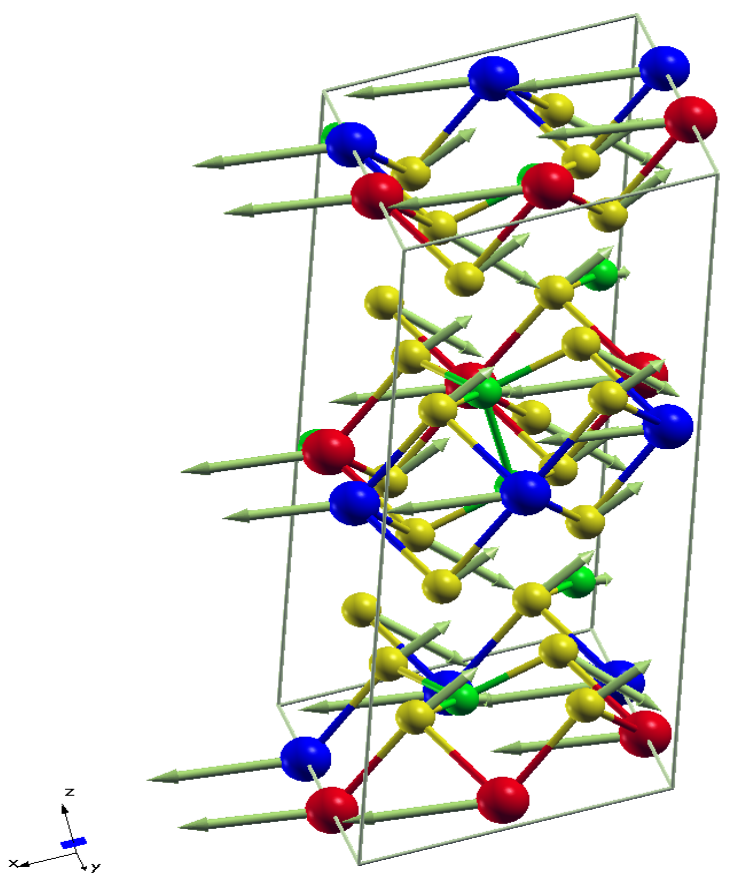}
         \caption{}
         \label{unstable3}
     \end{subfigure}
     \hfill
     \begin{subfigure}[b]{0.4\textwidth}
         \centering
         \includegraphics[width=\textwidth]{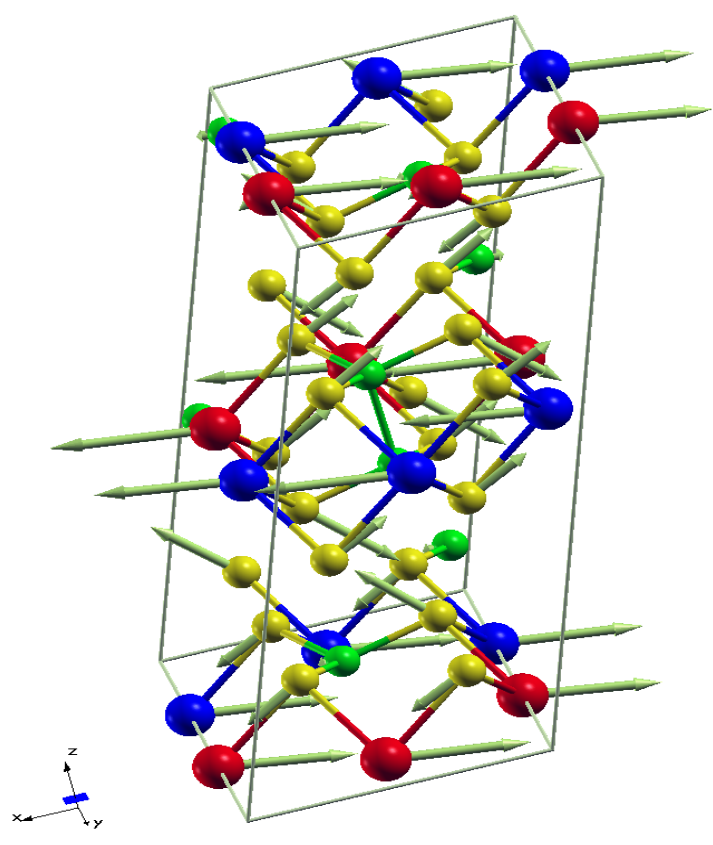}
         \caption{}
         \label{unstable4}
     \end{subfigure}
        \caption{\small{Unstable phonon modes of $C2/m$ (LP) phase at 0 GPa, (a) 153 \textit{i cm}$^{-1}$, (b) 151 \textit{i cm}$^{-1}$, (c) 143 \textit{i cm}$^{-1}$ and (d) 139 \textit{i cm}$^{-1}$. Red and blue atoms denote up-spin and down-spin Fe atoms, while green and yellow atoms are P and S atoms respectively. The most unstable mode gives the lowest energy structure for which no $\Gamma$-point instability was observed. Unstable modes were visualized using XCRYSDEN software \cite{xcrysden}.}}
        \label{LP_unstable}
\end{figure}

\begin{figure}[H]
     \begin{subfigure}[b]{0.3\textwidth}
         \centering
         \includegraphics[width=\textwidth]{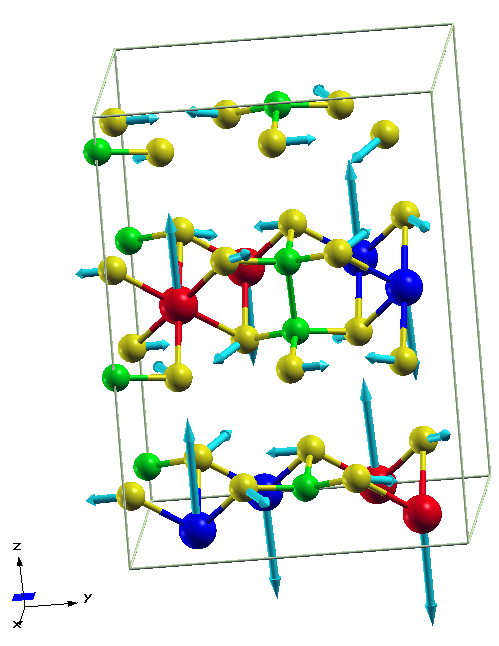}
        \caption{}
         \label{unstable1}
     \end{subfigure}
     \begin{subfigure}[b]{0.3\textwidth}
         \centering
        \includegraphics[width=\textwidth]{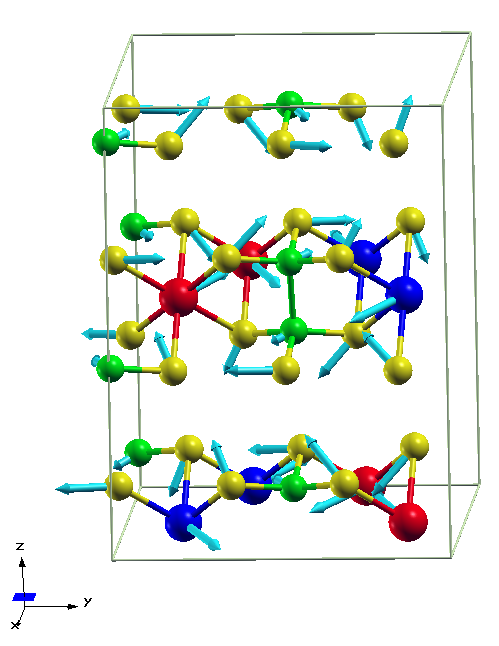}
         \caption{}
         \label{unstable2}
     \end{subfigure}
     \begin{subfigure}[b]{0.3\textwidth}
        \centering
         \includegraphics[width=\textwidth]{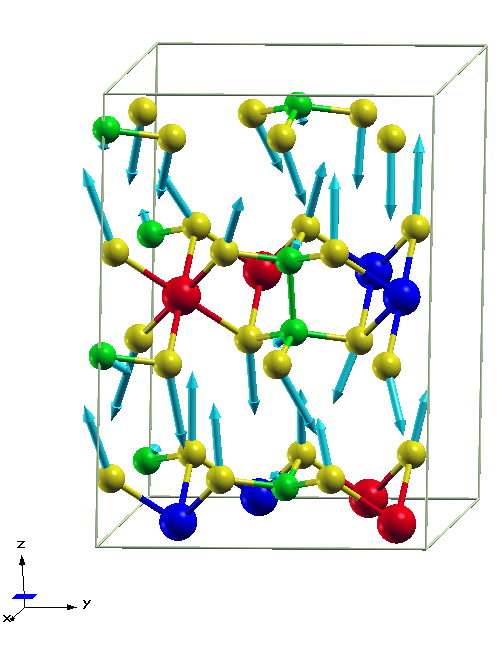}
         \caption{}
         \label{unstable3}
     \end{subfigure}
     \begin{subfigure}[b]{0.3\textwidth}
         \centering
         \includegraphics[width=\textwidth]{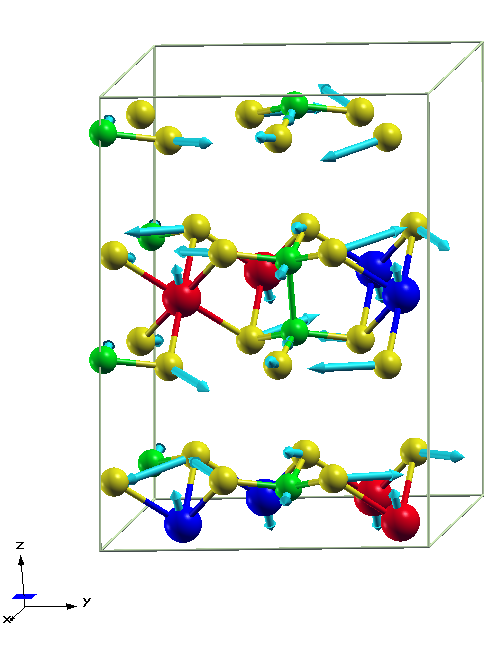}
         \caption{}
         \label{unstable4}
     \end{subfigure}
     \begin{subfigure}[b]{0.3\textwidth}
         \centering
         \includegraphics[width=\textwidth]{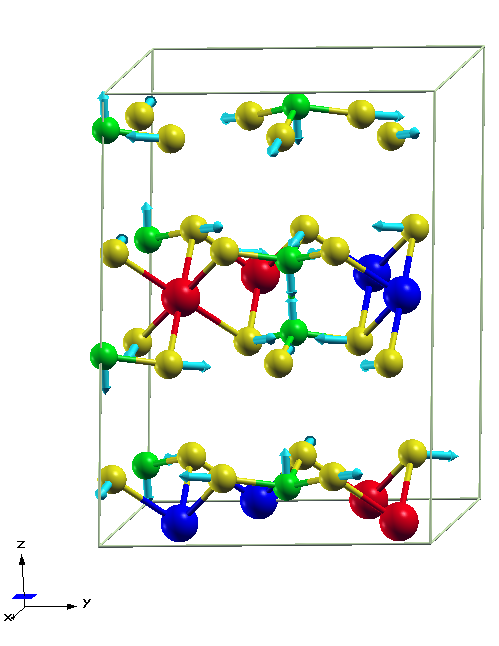}
         \caption{}
         \label{unstable4}
     \end{subfigure}
        \caption{\small{$\Gamma$-point Raman active modes of stable $C2/m$ (LP) phase (a) 93.6 \textit{cm}$^{-1}$ ($A_g, B_g$), (b) 154.4 \textit{cm}$^{-1}$ ($A_g$) (c) 239.2 \textit{cm}$^{-1}$  ($B_g$), (d) 269 \textit{cm}$^{-1}$ ($A_g, B_g$) and (e) 364.5 \textit{cm}$^{-1}$ ($A_g$) . $B_g$ mode at 239.2 \textit{cm}$^{-1}$ has highest $d\omega/dp$ of $\sim$5 $cm^{-1}GPa^{-1}$ which involves vibration of S atoms against each other in individual monolayers along layered direction i.e c-direction. Red and blue atoms denote up-spin and down-spin Fe atoms, while green and yellow atoms are P and S atoms respectively. Modes were visualized using XCRYSDEN software \cite{xcrysden}.}}
        \label{LP_modes}
\end{figure}
\begin{figure}[H]
\centering
     \begin{subfigure}[b]{0.31\textwidth}
         \centering
         \includegraphics[width=\textwidth]{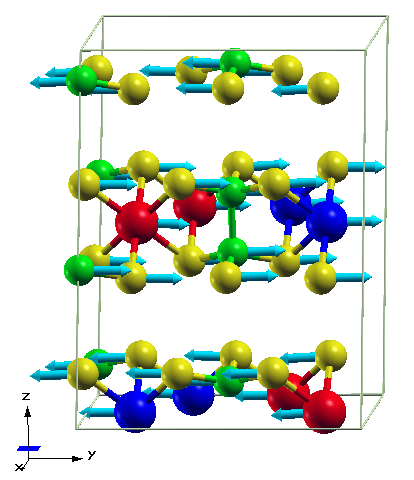}
        \caption{}
         \label{soft1}
     \end{subfigure}
     \begin{subfigure}[b]{0.26\textwidth}
         \centering
        \includegraphics[width=\textwidth]{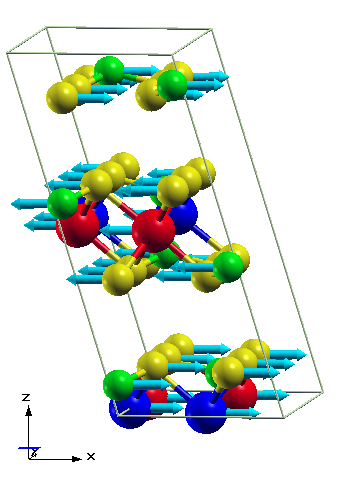}
         \caption{}
         \label{soft2}
     \end{subfigure}
     \begin{subfigure}[b]{0.31\textwidth}
        \centering
         \includegraphics[width=\textwidth]{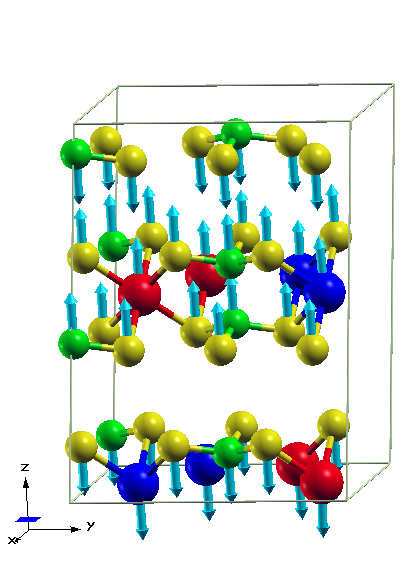}
         \caption{}
         \label{soft3}
     \end{subfigure}
         \begin{subfigure}[b]{0.5\textwidth}
        \centering
         \includegraphics[width=\textwidth]{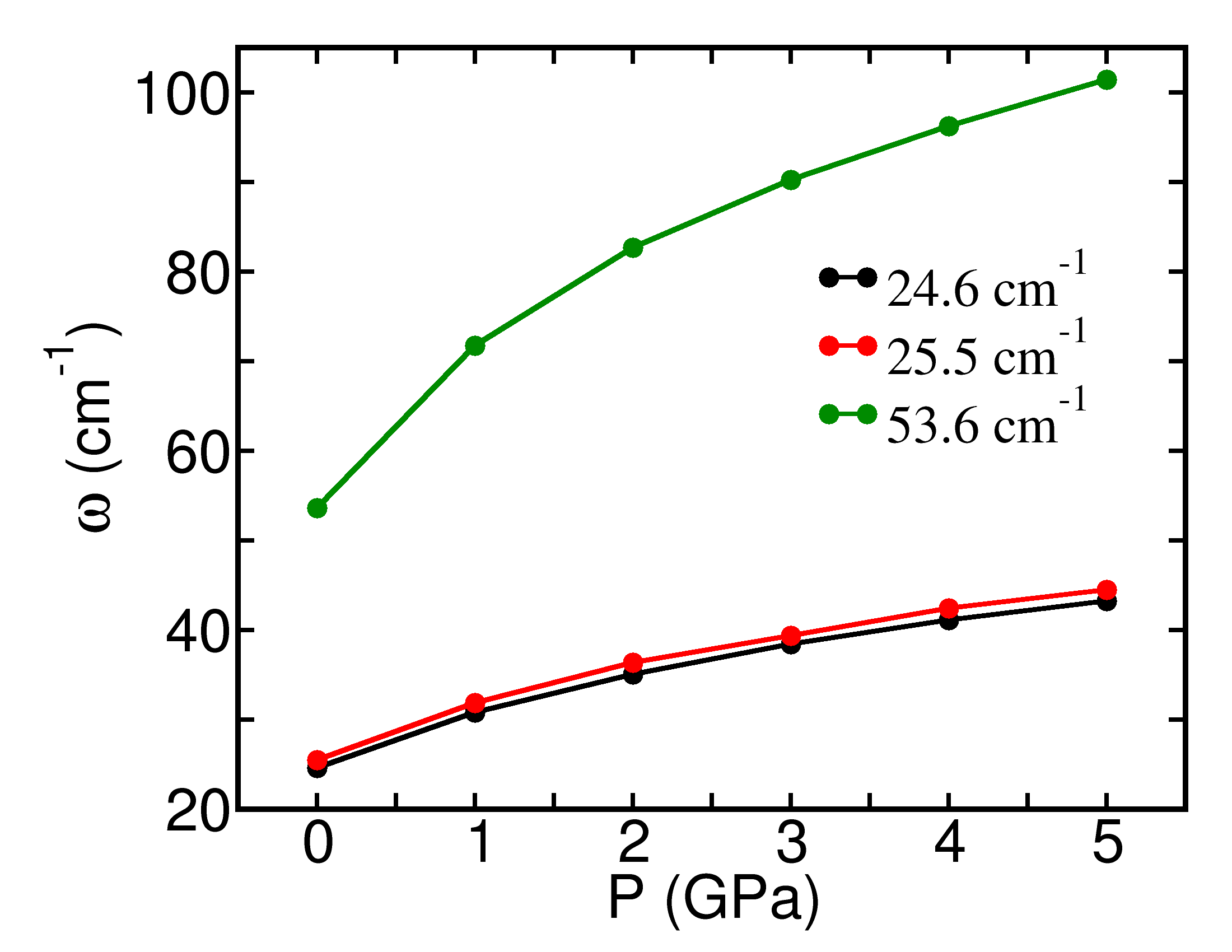}
         \caption{}
         \label{plot}
     \end{subfigure}
        \caption{\small{Soft modes in stable $C2/m$ (LP) phase at $\Gamma$-point (a) 24.6 \textit{cm}$^{-1}$ ($B_g$), (b) 25.5 \textit{cm}$^{-1}$ ($B_g$) and (c) 53.6 \textit{cm}$^{-1}$  ($B_g$). Mode (c), where individual monolayers vibrate against each other along $c$-direction, shows an increase of $\sim$48 \textit{cm}$^{-1}$ from 0 to 5 GPa and increases rapidly with pressure as compared to modes (a) and (b). Slope of mode (c) gradually decreases near 5GPa, which can be due to increasing repulsion between individual monolayers at high pressures. Mode (a) and (b), where monolayers slide against each other along $b$ and $a$-direction, show an increase of $\sim$18 \textit{cm}$^{-1}$ from 0 to 5 GPa. All three modes deviate from linear behavior and harden with pressure. Red and blue atoms denote up-spin and down-spin Fe atoms, while green and yellow atoms are P and S atoms respectively. Modes were visualized using XCRYSDEN software \cite{xcrysden}.}}
        \label{LP_soft_modes}
\end{figure}
\begin{figure}[H]
\centering
      \includegraphics[width=0.7\textwidth,clip]{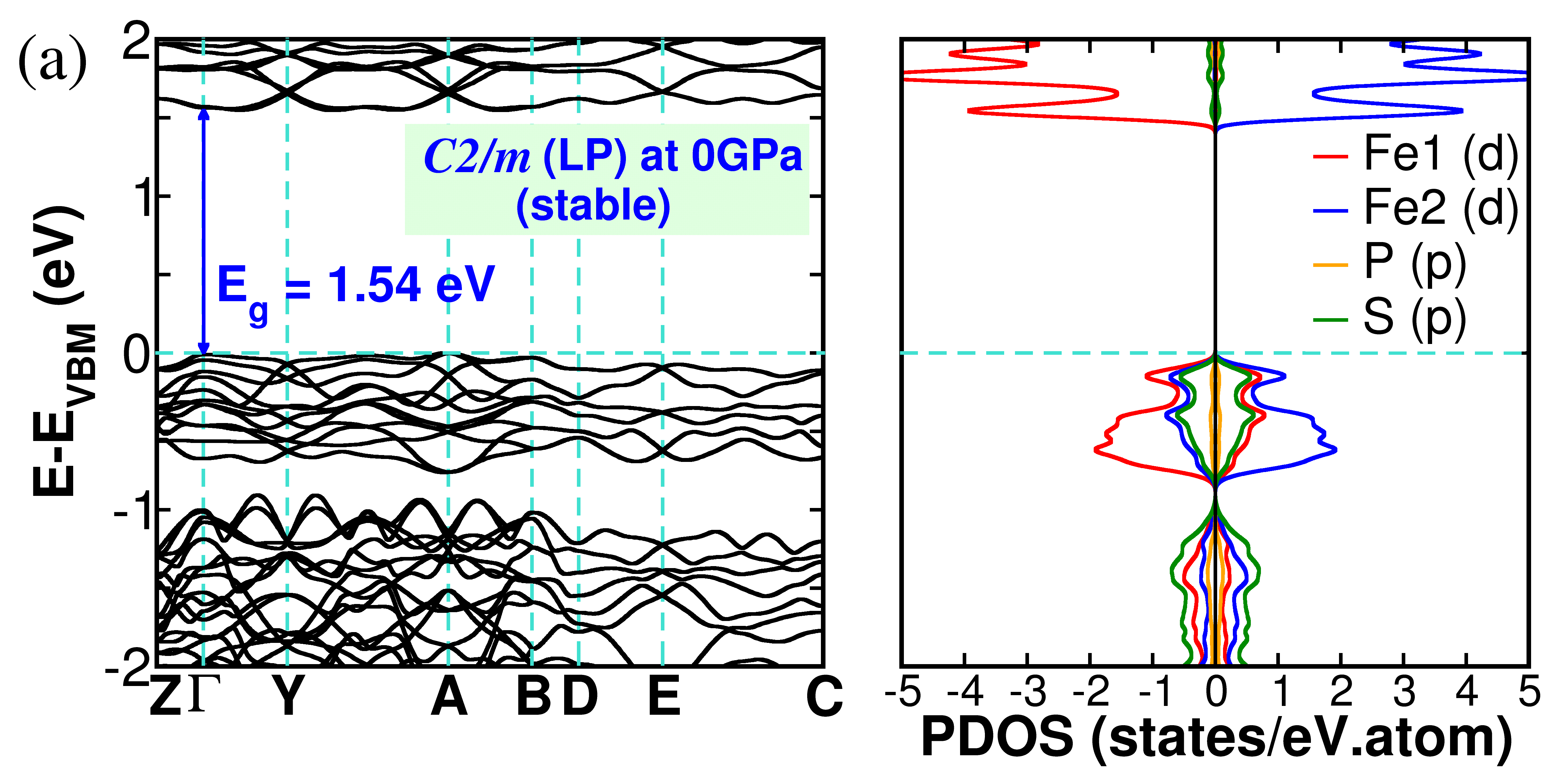}
     \includegraphics[width=0.7\textwidth,clip]{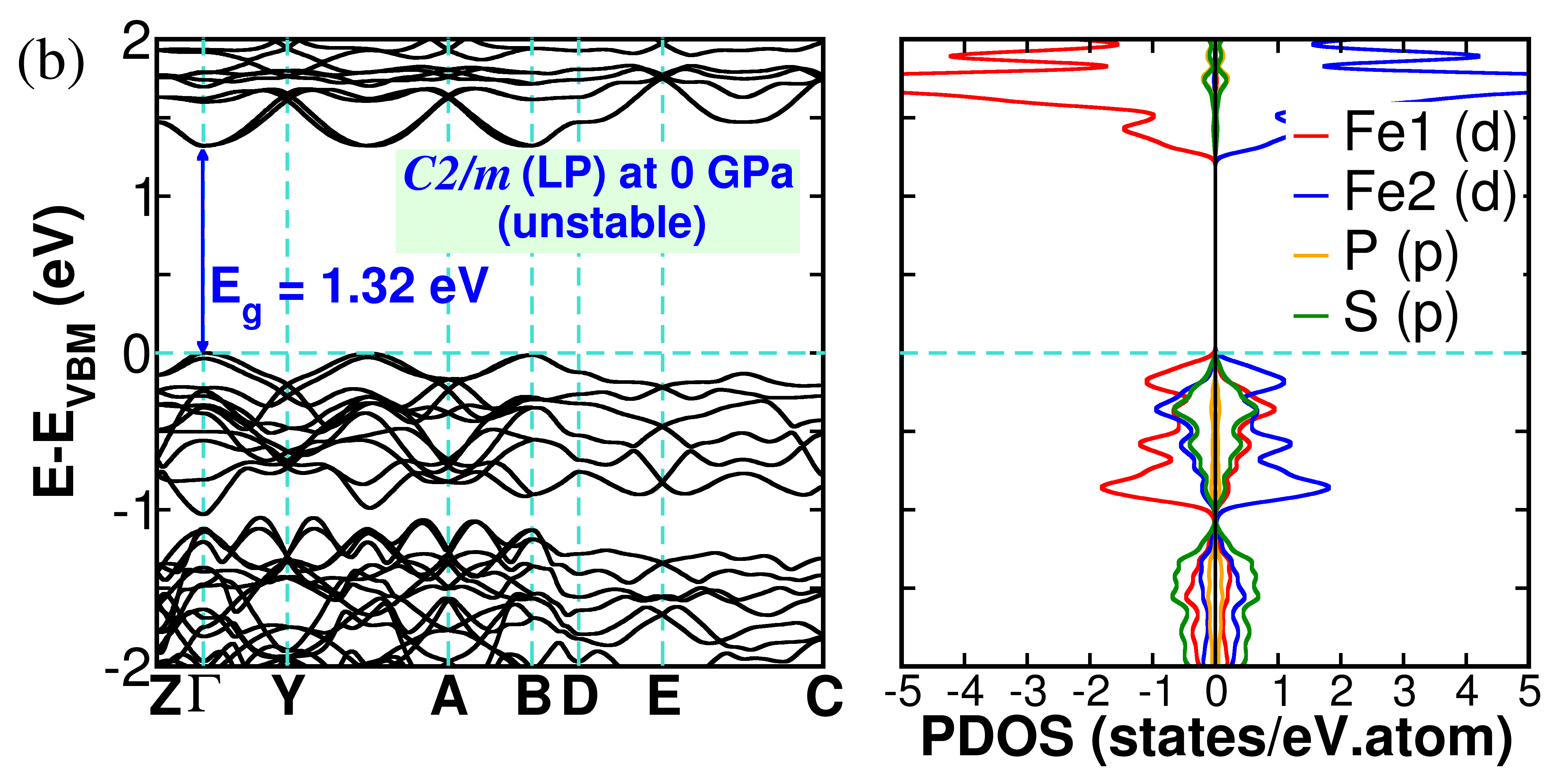}
\caption{\small{Electronic structure of $C2/m$ (LP) phase FePS$_3$ for (a) stable structure, without any phonon instability at $\Gamma$-point , and (b) unstable structure with most unstable mode at $153 i $ cm$^{-1}$. A band gap of 1.32 eV was observed for dynamically unstable structure. The stable structure has a larger band gap of 1.54 eV. PDOS shows changes in contribution of orbitals of different atoms.}}
\label{c2m_lp_bands}
\end{figure}

\begin{figure}[H]
     \begin{subfigure}[b]{0.29\textwidth}
         \centering
         \includegraphics[width=\textwidth]{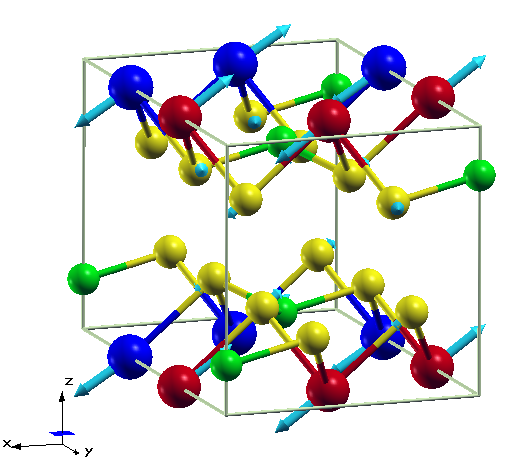}
        \caption{}
         \label{unstable1}
     \end{subfigure}
     \begin{subfigure}[b]{0.32\textwidth}
         \centering
        \includegraphics[width=\textwidth]{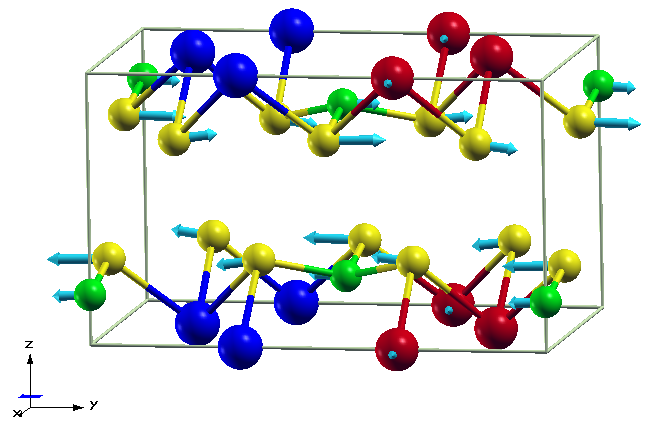}
         \caption{}
         \label{unstable2}
     \end{subfigure}
     \begin{subfigure}[b]{0.32\textwidth}
        \centering
         \includegraphics[width=\textwidth]{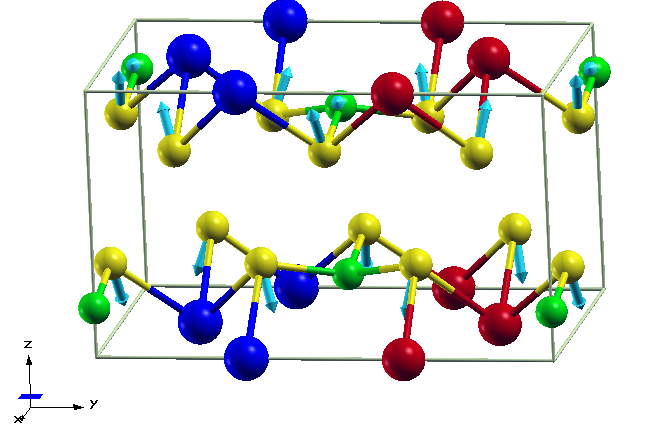}
         \caption{}
         \label{unstable3}
     \end{subfigure}
     \begin{subfigure}[b]{0.32\textwidth}
         \centering
         \includegraphics[width=\textwidth]{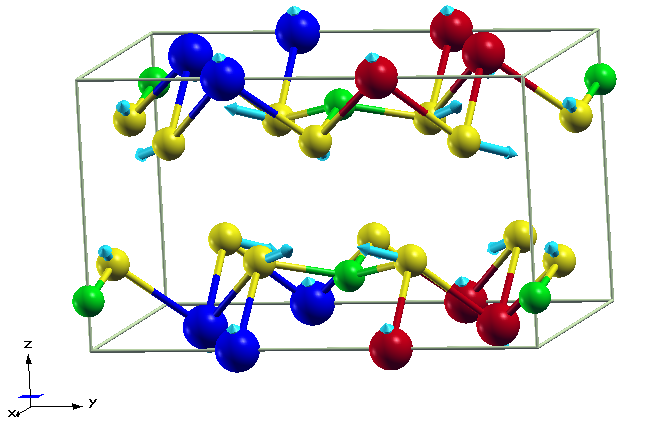}
         \caption{}
         \label{unstable4}
     \end{subfigure}
     \begin{subfigure}[b]{0.32\textwidth}
         \centering
         \includegraphics[width=\textwidth]{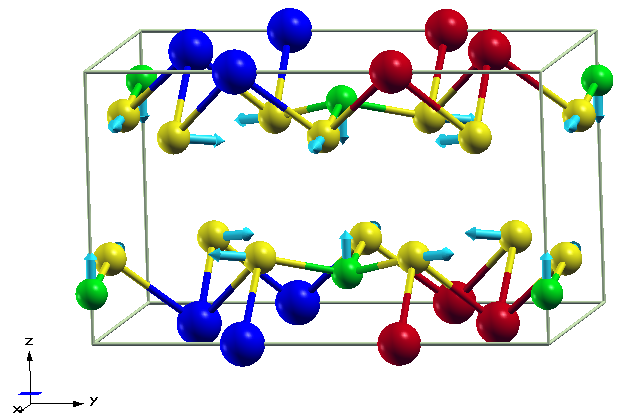}
         \caption{}
         \label{unstable4}
     \end{subfigure}
        \caption{\small{$\Gamma$-point modes of  $C2/m$ (HP-I) phase at 6 GPa (a) 99.1 \textit{cm}$^{-1}$ ($ B_g$), (b) 171.8 \textit{cm}$^{-1}$ ($B_g$), (c) 270.6 \textit{cm}$^{-1}$  ($A_g$), (d) 289.3 \textit{cm}$^{-1}$ ($B_g$) and (e) 383.2 \textit{cm}$^{-1}$ ($A_g$) . Red and blue atoms denote up-spin and down-spin Fe atoms, while green and yellow atoms are P and S atoms respectively. Modes were visualized using XCRYSDEN software \cite{xcrysden}.}}
        \label{HP_modes}
\end{figure}
\begin{figure}[H]
     \begin{subfigure}[b]{0.34\textwidth}
         \centering
         \includegraphics[width=\textwidth]{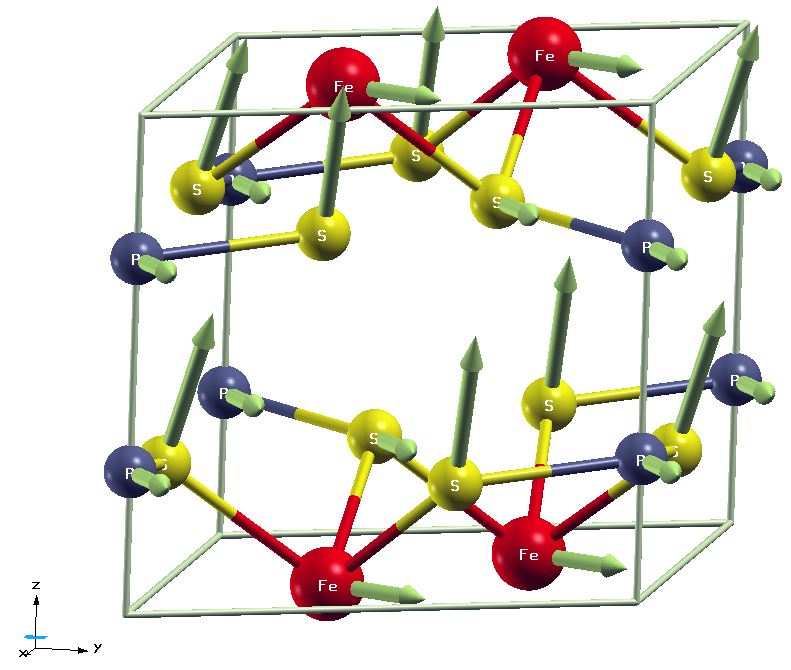}
        \caption{M$_1$ ($E_g$)}
         \label{eg1}
     \end{subfigure}
     \hfill
     \begin{subfigure}[b]{0.34\textwidth}
         \centering
        \includegraphics[width=\textwidth]{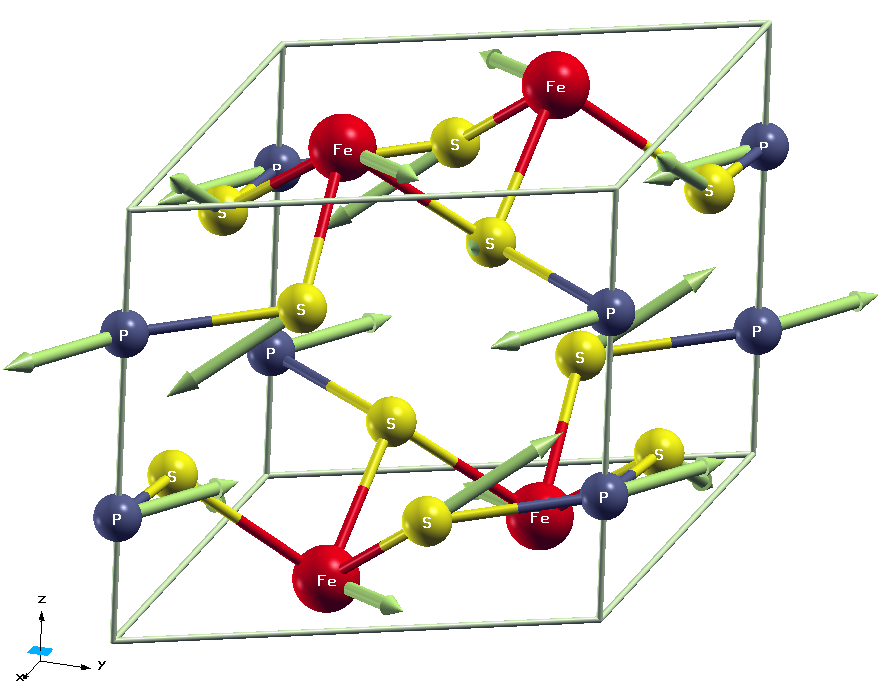}
         \caption{M$_2$ ($E_g$)}
         \label{eg2}
     \end{subfigure}
     \hfill
     \begin{subfigure}[b]{0.34\textwidth}
        \centering
         \includegraphics[width=\textwidth]{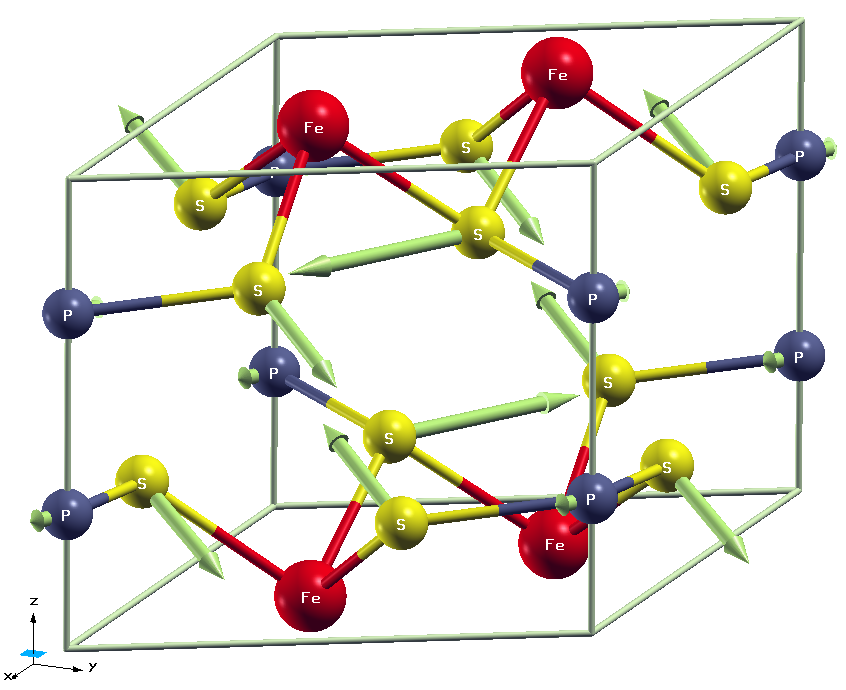}
         \caption{M$_3$ ($E_g$)}
         \label{eg3}
     \end{subfigure}
     \hfill
     \begin{subfigure}[b]{0.34\textwidth}
         \centering
         \includegraphics[width=\textwidth]{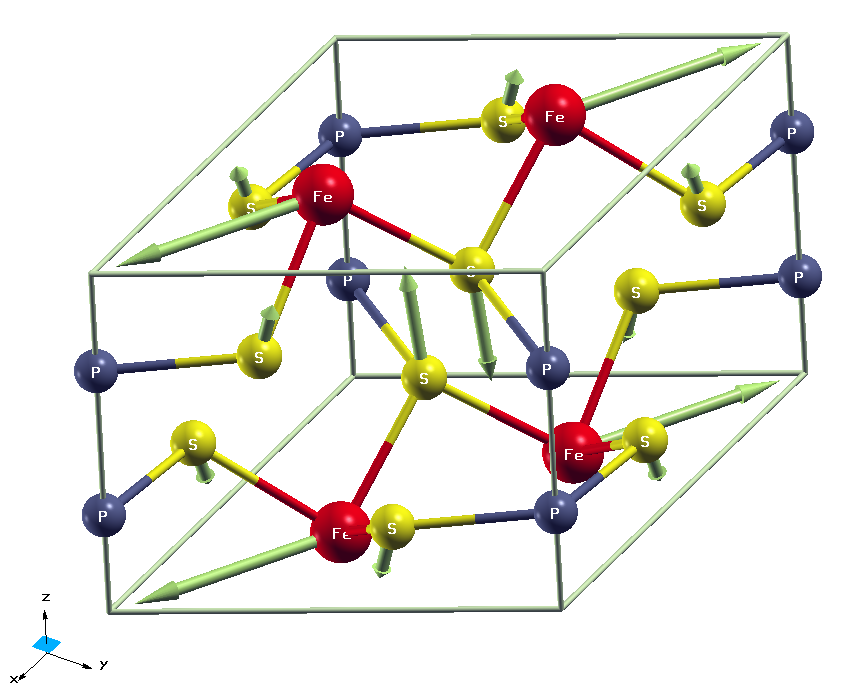}
         \caption{M$_5$ ($E_g$)}
         \label{eg4}
     \end{subfigure}
     \begin{subfigure}[b]{0.34\textwidth}
        \centering
         \includegraphics[width=\textwidth]{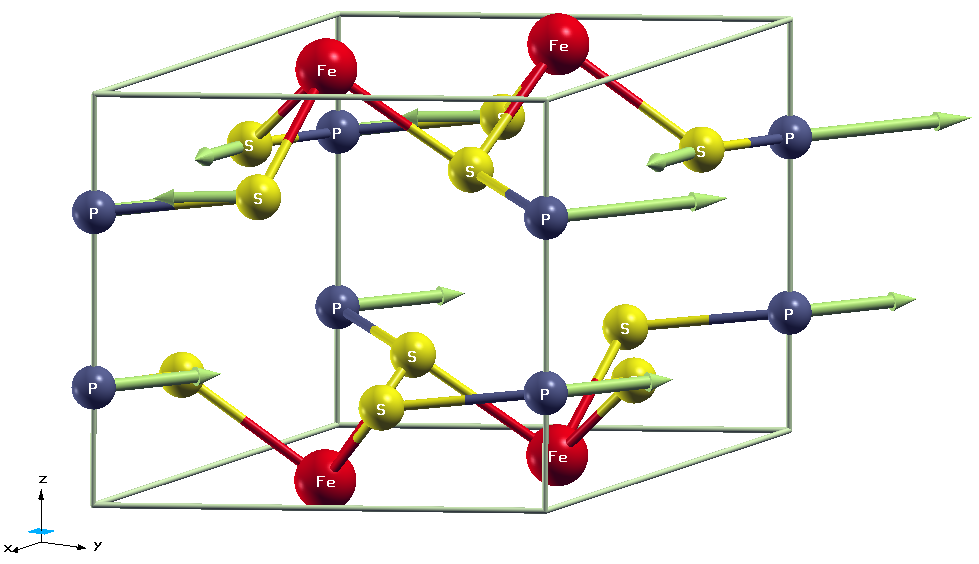}
         \caption{M$_8$ ($E_g$)}
         \label{eg5}
     \end{subfigure}
     \hfill
     \begin{subfigure}[b]{0.34\textwidth}
        \centering
         \includegraphics[width=\textwidth]{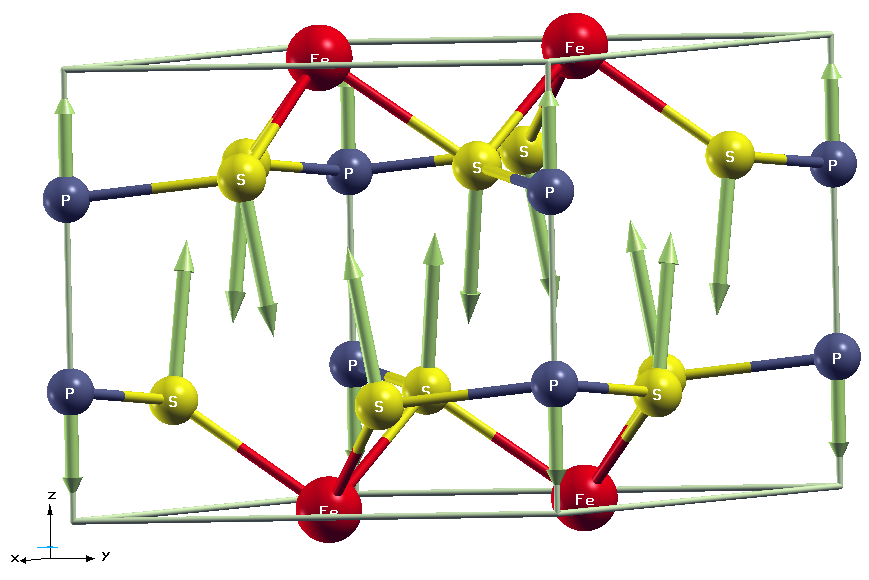}
         \caption{M$_4$ ($A_{1g}$)}
         \label{a1g1}
     \end{subfigure}
     \hfill
     \begin{subfigure}[b]{0.34\textwidth}
        \centering
         \includegraphics[width=\textwidth]{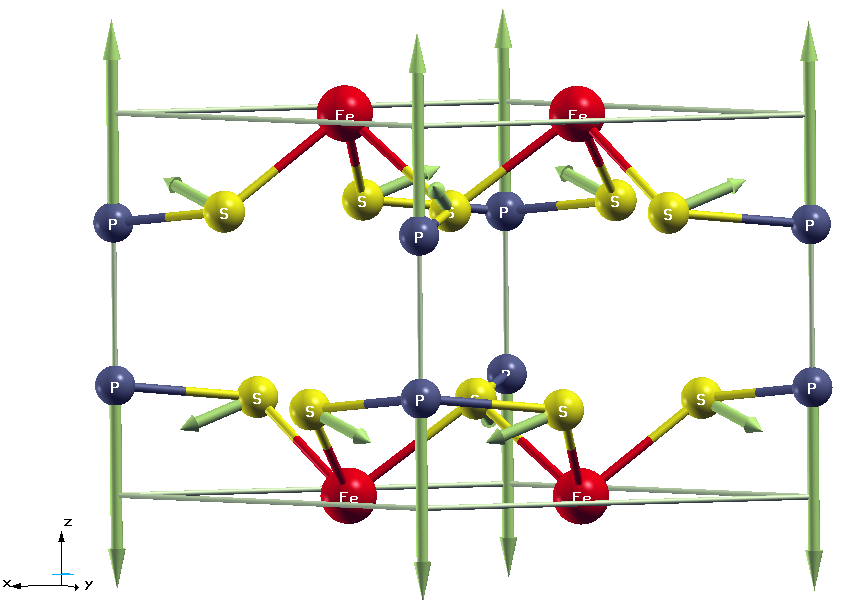}
         \caption{M$_6$ ($A_{1g}$)}
         \label{a1g2}
     \end{subfigure}
     \hfill
     \begin{subfigure}[b]{0.34\textwidth}
        \centering
         \includegraphics[width=\textwidth]{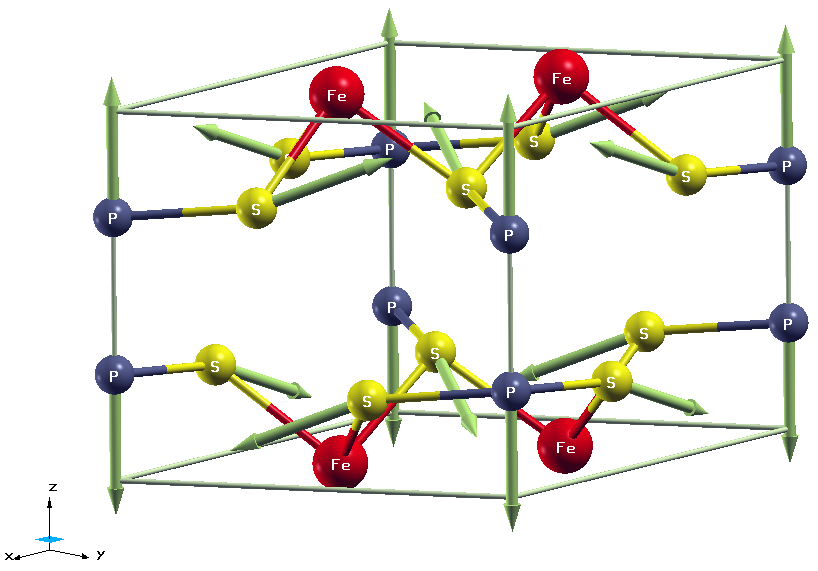}
         \caption{M$_7$ ($A_{1g}$)}
         \label{a1g3}
     \end{subfigure}
     \hfill
        \caption{\small{$\Gamma$-point modes of HP-II phase. Red, blue and yellow atoms are Fe, P and S atoms respectively. Modes from (a) to (e) are $E_g$ and (f) to (h) are $A_{1g}$ modes, which are given in increasing order of frequency. Modes were visualized using XCRYSDEN software \cite{xcrysden}.}}
        \label{p31m_modes}
\end{figure}

\begin{figure}[H]
        \centering
         \includegraphics[width=1.3\textwidth,angle=90]{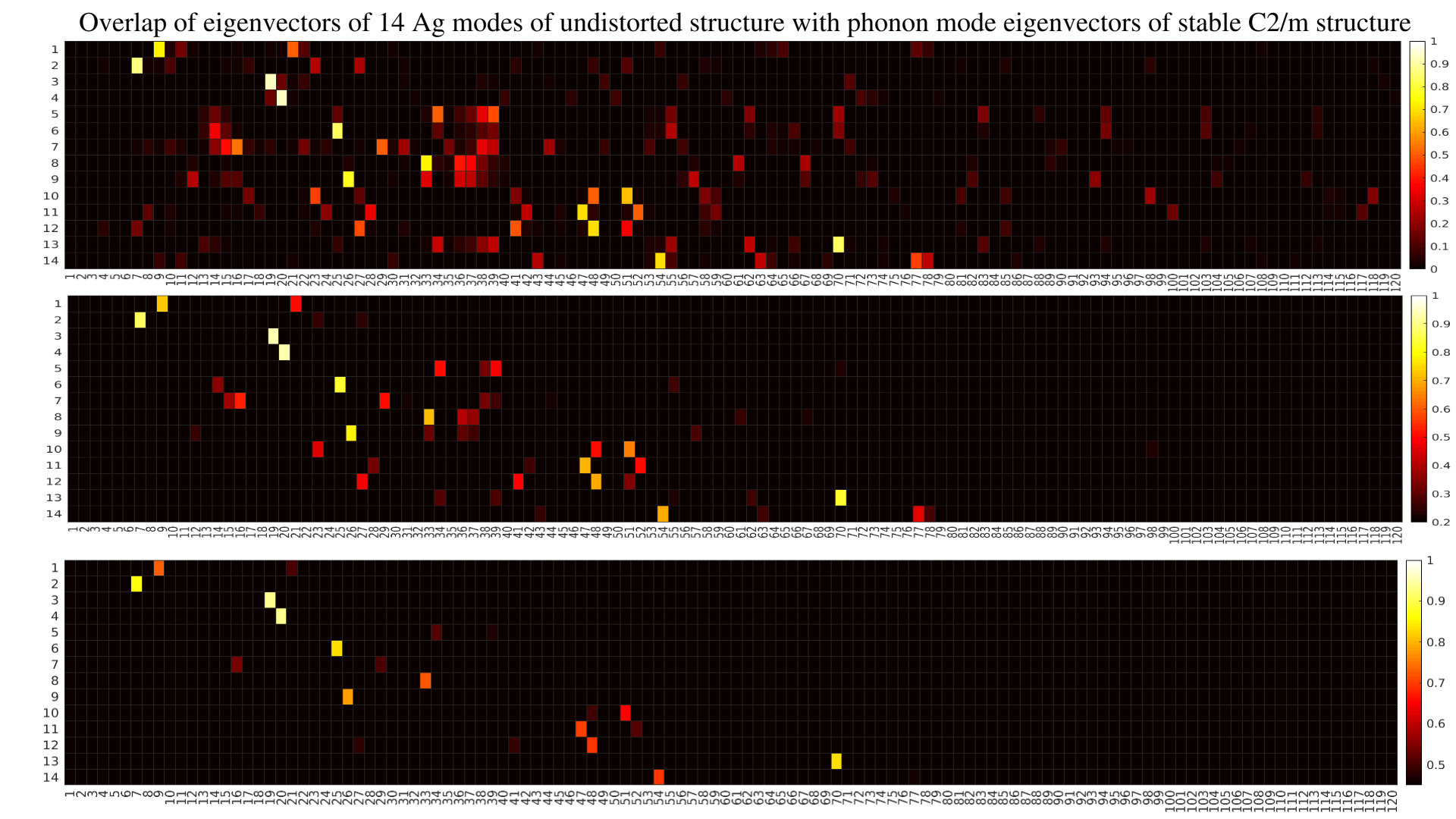}
\caption{Overlap of phonon mode eigenvector of $A_g$ modes of undistorted structure with phonon mode eigenvector of stable C2/m structure, obtained after displacing atoms along unstable mode and relaxing the structure.}
\end{figure}

\begin{figure}[H]
        \centering
         \includegraphics[width=1.3\textwidth,angle=90]{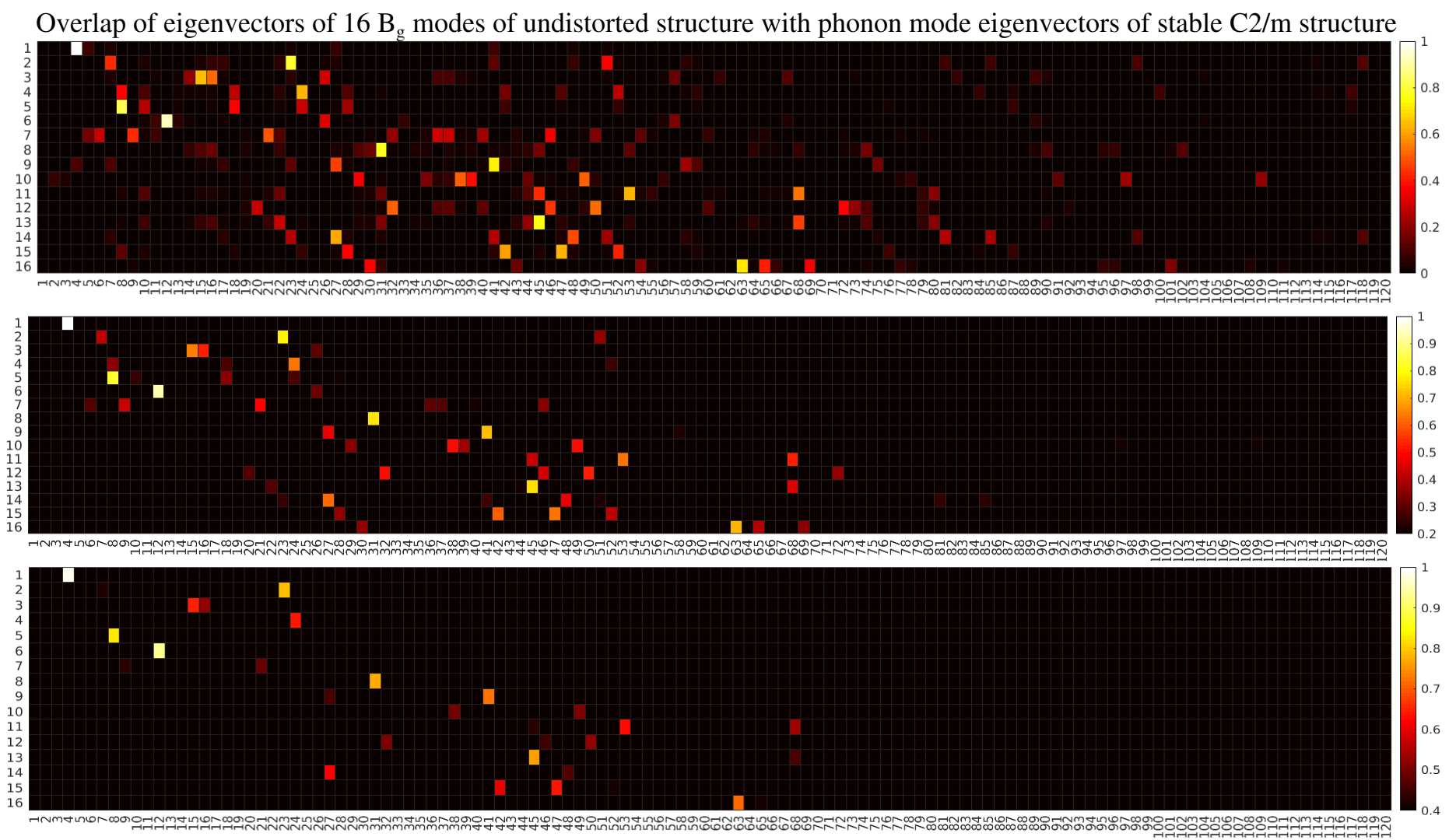}
\caption{Overlap of phonon mode eigenvector of $B_g$ modes of undistorted structure with phonon mode eigenvector of stable C2/m structure.}
\end{figure}
\clearpage
\bibliographystyle{apsrev4-2}
\bibliography{ref2}